\documentclass[12pt]{article}

\pdfoutput=1

\usepackage{latexsym}
\usepackage{epsfig,amssymb,euscript}
\usepackage{amsmath}
\usepackage{mathrsfs}
\numberwithin{equation}{section}  
\newsavebox{\ns}
\newsavebox{\dbrane}

\def\be{\begin{equation}}
\def\ee{\end{equation}}
\def\bea{\begin{eqnarray}}
\def\eea{\end{eqnarray}}

\renewcommand{\theequation}{\arabic{section}.\arabic{equation}}
\def\theequation{\thesection.\arabic{equation}}

\def\Dslash{\,\,{\raise.15ex\hbox{/}\mkern-12mu D}}
\def\Dbarslash{\,\,{\raise.15ex\hbox{/}\mkern-12mu {\bar D}}}
\def\delslash{\,\,{\raise.15ex\hbox{/}\mkern-9mu \partial}}
\def\delbarslash{\,\,{\raise.15ex\hbox{/}\mkern-9mu {\bar\partial}}}
\def\pslash{\,\,{\raise.15ex\hbox{/}\mkern-9mu p}}
\def\calDslash{\,\,{\raise.15ex\hbox{/}\mkern-12mu {\cal D}}}

\newcommand{\vol}{\mathrm{vol}}
\newcommand{\nn}{\nonumber \\}
\newcommand{\reef}[1]{(\ref{#1})}

\newcommand{\bbZ}{\mathbb{Z}}
\newcommand{\bbR}{\mathbb{R}}

\newcommand{\hfac}{(16\pi G/3)^{1/2}}
\newcommand{\mfac}{(16\pi G)^{1/2}}

\begin{document}

\makeatletter
\renewcommand{\theequation}{\thesection.\arabic{equation}}
\@addtoreset{equation}{section}
\makeatother

\baselineskip 18pt

\begin{titlepage}

\vfill

\begin{flushright}
Imperial/TP/2009/JG/08\\
\end{flushright}

\vfill

\begin{center}
   \baselineskip=16pt
   {\Large\bf  Quantum Criticality and Holographic Superconductors in M-theory}
  \vskip 1.5cm
      Jerome P. Gauntlett, Julian Sonner and Toby Wiseman\\
   \vskip .6cm
  \begin{small}
      \textit{Theoretical Physics Group, Blackett Laboratory, \\
        Imperial College, Prince Consort Rd, London SW7 2AZ, U.K.}
        \end{small}\\*[.6cm]
   \end{center}

\vfill

\begin{center}
\textbf{Abstract}
\end{center}

\begin{quote}
We present a consistent Kaluza-Klein truncation of $D=11$ supergravity on an arbitrary
seven-dimensional Sasaki-Einstein space ($SE_7$) to a $D=4$ theory containing a metric, a 
gauge-field, a complex scalar field and a real scalar field. We use this $D=4$ theory to construct 
various black hole solutions that describe the thermodynamics of the $d=3$ CFTs dual to skew-whiffed $AdS_4\times SE_7$ solutions.
We show that these CFTs have a rich phase diagram, including holographic superconductivity
with, generically, broken parity and time reversal invariance. At zero temperature the superconducting solutions 
are charged domain walls with a universal emergent conformal symmetry in the far infrared.

\end{quote}

\vfill

\end{titlepage}
\setcounter{equation}{0}

\tableofcontents
\section{Introduction}
Black holes with charged hair in anti-de-Sitter space provide a
holographic description of superconductivity via the AdS/CFT correspondence.
This idea was first discussed in \cite{Gubser:2008px}\cite{H31}
and some of the subsequent developments have been
nicely reviewed in \cite{Hartnoll:2009sz}\cite{Herzog:2009xv}.
In more detail, one considers a CFT with an $AdS$ dual in a theory of gravity
with matter fields which include a Maxwell gauge field and additional charged fields.
The CFT is studied at finite temperature $T$ and at fixed chemical potential
$\mu$ (or at fixed charge density) by studying electrically charged black holes in the dual gravity theory.
Charged black holes with vanishing charged matter fields describe the
high temperature, normal phase of the superconductor. Below
some critical temperature, one requires a new
branch of charged black hole solutions to emerge
that carry charged hair and are thermodynamically favoured.
The charged hair spontaneously breaks the $U(1)$ gauge symmetry in the bulk which  corresponds to a spontaneous breaking
of a global $U(1)$ symmetry in the boundary CFT, signalling the superconductivity. More precisely, this signals superfluidity,
but for certain phenomena one expects that the difference between the two is not significant \cite{H32}.

Most work has focussed on $D=4$ theories of gravity, corresponding to superconductors in $d=3$ spacetime dimensions,
as this is likely to be the most promising arena to make contact with real materials.
The black holes are usually taken to have flat $\bbR^2$ horizons, and hence are also called black branes,
corresponding to considering the CFTs in $d=3$ Minkowski spacetime. The conformal invariance then implies
that the critical temperature for the onset of superconductivity is fixed by the scale set by $\mu$.

Most studies have been carried out within the context of ``phenomenological'' models of gravity without any
obvious embedding into string/M-theory. This bottom up approach has the virtue of simplicity but has the drawback that
one is not guaranteed that there is a well defined
underlying conformal field theory. Furthermore, if the models are posited to
just provide approximate supergravity solutions to string/M-theory, the approximations can obscure important physical features,
such as the low-temperature behaviour of the superconductors.
Many investigations have also used a ``probe approximation'' within these phenomenological
models, in which the back reaction on the gravitational field is ignored. This approximation again makes the analysis
more tractable and while it should capture some important features it cannot be used to study, for example, low-temperature phenomena.
The incorporation of back reaction for the most studied class of phenomenological models with a single charged scalar field was initiated in
the foundational work \cite{H32}.

There has also been recent progress in a top down approach, where one aims to find exact solutions of
$D=11$ and type IIB supergravity. The constructions have been based on new consistent KK truncations of the
supergravity theories.
Building on the work of \cite{dh}\cite{Gauntlett:2009zw}, fully back reacted solutions
of $D=11$ supergravity that describe holographic superconductors in $d=3$ spacetime dimensions
were constructed in \cite{Gauntlett:2009dn}. In the zero temperature limit, $T\to 0$, the event horizon of the
superconducting black holes disappears implying that the entropy of the dual superconductors vanishes in this
limit. More precisely, it was shown in \cite{Gauntlett:2009dn} (and further studied in \cite{Gubser:2009gp})
that as $T\to 0$ the superconducting black hole solutions
approach charged domain wall solutions that interpolate between (perturbed) $AdS_4$ solutions in the UV that
preserve the $U(1)$ gauge symmetry and different $AdS_4$ solutions in the IR that break the gauge symmetry.
These domain wall solutions demonstrate that when the dual CFT is held at zero temperature
and finite chemical potential there is an emergent conformal symmetry at low energies,
exactly as in a class of phenomenological models\footnote{Note that models with an interesting emergent Lifshitz scaling in
the IR were also studied in \cite{Gubser:2009cg}.} studied in \cite{Gubser:2008wz}.
The $D=11$ superconducting black hole solutions of \cite{Gauntlett:2009dn} were constructed using a $D=4$ theory
of gravity, to be discussed momentarily. This $D=4$ theory has some similarities with the $D=4$ phenomenological model of \cite{H32} after fixing
some parameters. However, there are also important differences. In particular, it was shown in \cite{Gauntlett:2009dn} that the superconducting black hole
solutions in the model of \cite{H32}, for these parameters\footnote{The $T\to 0$ limit for other values of the parameters in the model of \cite{H32}
was discussed in \cite{hr}.},
become singular as $T\to 0$.

Analogous constructions of solutions of type IIB supergravity dual to superconductors
in $d=4$ spacetime dimensions have been carried out in \cite{Gubser:2009qm}\cite{Gubser:2009gp}, but not
yet in the same level of detail. It has been shown that superconducting black hole solutions
should exist using a probe analysis in \cite{Gubser:2009qm}. Furthermore, zero temperature domain wall
solutions with emergent conformal symmetry have been constructed \cite{Gubser:2009gp}; while it is expected that they arise
as the zero temperature limit of the superconducting black hole solutions at finite temperature, this has not yet been shown.
These superconducting black hole solutions in $D=11$ and type IIB carry abelian charge of ``$R$-symmetry type'' (it is an $R$-symmetry for supersymmetric vacua); 
the construction of type IIB solutions carrying a baryonic abelian charge were recently presented in
\cite{Herzog:2009gd}.

In this paper we will expand upon and extend the analysis of $d=3$ holographic superconductivity arising in $D=11$ supergravity
that was initiated in \cite{Gauntlett:2009dn}.
The solutions found in \cite{Gauntlett:2009dn} were obtained
using the consistent KK truncation \cite{Gauntlett:2009zw}
of $D=11$ supergravity on a seven-dimensional
Sasaki-Einstein space down to a $D=4$ theory of gravity that includes six real scalar fields and two
gauge fields.
The consistency of this KK reduction means that given an arbitrary Sasaki-Einstein metric,
any solution of the $D=4$ theory can be uplifted to obtain an exact solution of $D=11$ supergravity.
It was shown in \cite{Gauntlett:2009dn} that it is possible to further truncate the $D=4$ theory to a theory with a metric, $g$,
a gauge field, $A$, with field strength $F$, and a charged scalar field, $\chi$, provided that one
restricts to solutions satisfying $F\wedge F=0$ (such as electrically charged black holes).
One result of this paper is that there is a consistent truncation, with no additional restrictions,
that is obtained by also keeping an additional neutral scalar field $h$.
As we now discuss this truncated $D=4$ theory with  fields $(g,A,\chi,h)$, with action given in \reef{lagred},
has a rich structure to explore
different aspects of superconductivity.
A particularly interesting feature of the $D=4$ theory is that the neutral scalar field $h$ couples to $F\wedge F$
and this coupling implies that the solutions we discuss with
$h\ne 0$ break $d=3$ parity and time reversal invariance.

The truncated theory has three $AdS_4$ vacua which uplift to $AdS_4\times SE_7$ solutions in $D=11$.
One of them uplifts to the skew-whiffed $AdS_4$ solutions of Freund-Rubin type \cite{Freund:1980xh}, as shown
in \cite{Gauntlett:2009zw},
the second uplifts to the $AdS_4$ solutions of
Pope and Warner type \cite{Pope:1984bd}\cite{Pope:1984jj},
as shown in \cite{Gauntlett:2009dn},
and the third, which is the only one with $h\ne 0$, uplifts to the $AdS_4$ solutions of Englert type \cite{Englert:1982vs}\cite{Awada:1982pk},
as we will show here. Recall that the skew-whiffed solutions do not preserve any supersymmetry, except in the special case
that the $SE_7$ space is the round seven-sphere in which case it preserves all supersymmetry. All
skew-whiffed solutions are known to be perturbatively stable \cite{Duff:1984sv}. Some discussion on
the possibility of $1/N$ effects destabilising the non-supersymmetric skew-whiffed
solutions has been discussed in \cite{Berkooz:1998qp}\cite{am}.
The Pope-Warner and the Englert solutions do not preserve any supersymmetry for any choice of $SE_7$.
A stability analysis for the Pope-Warner solutions has not been carried out; in light of the results presented
in \cite{Gauntlett:2009dn} and here, we feel that this would now be a worthwhile investigation.
On the other hand the Englert solutions are known to be unstable \cite{Page:1984fu}:
indeed we will show that an unstable mode is already present in our truncated Lagrangian.

In the skew-whiffed $AdS_4$ vacuum the operators ${\cal O}_\chi, {\cal O}_h$ dual to the fields $\chi,h$, respectively,
are both relevant operators with scaling dimension taken\footnote{This arises because of the boundary conditions we shall impose on the
skew-whiffed $AdS_4$ solution; different boundary conditions would lead to $\Delta=1$.} to be $\Delta=2$.
Before investigating superconductivity, we first construct uncharged domain wall solutions (i.e. the gauge-field is identically zero)
that describe ordinary RG flows between the different $AdS$ vacua. We will show that there is a one parameter family of such
domain wall solutions
which describe RG flows between the skew-whiffed vacuum in the UV, perturbed by
${\cal O}_\chi$ and ${\cal O}_h$,
and the Pope-Warner vacuum in the IR. As is usual for such RG flows the operators have non-zero vevs, $<{\cal O}_\chi>,<{\cal O}_h>\ne 0$.
We will also show that there is a single domain wall solution that
flows between the skew-whiffed vacuum in the UV to the Englert vacuum in the IR and also a domain wall solution that
flows from the Englert vacuum in the UV to the Pope-Warner vacuum in the IR. Of course, given the instability of the Englert vacuum,
the physical relevance of these latter domain wall solutions is not clear.

We then turn our attention to superconductivity. We will be interested in the CFT dual to the skew whiffed vacuum that has been
deformed by ${\cal O}_h$, generalising the analysis of \cite{Gauntlett:2009dn},
and the solutions we construct imply that we also have $<{\cal O}_h>\ne 0$. The superconductivity is signalled by a spontaneous breaking of the $U(1)$ symmetry, so we demand that the CFT is not perturbed by ${\cal O}_\chi$  (in contrast to the RG flow solutions discussed in the last paragraph) and look for solutions with $<{\cal O}_\chi>\ne 0$.

We first construct charged domain walls that describe the deformed skew-whiffed CFT
at zero temperature and non-zero chemical potential. After observing that in the Pope-Warner vacuum the gauge-field is dual to an irrelevant operator
a simple parameter count suggests that, given the one parameter family of uncharged domain wall RG flow solutions mentioned above,
there should also be a one parameter family of charged domain wall solutions that interpolate between the skew-whiffed vacuum in the UV and the Pope-Warner vacuum in the IR.
This is in accord with the conjecture made in \cite{Gubser:2009gp}. In the special case that $h=0$ such a solution was found in \cite{Gauntlett:2009dn} (and further studied in\cite{Gubser:2009gp})
and here we will construct a more general one parameter family of solutions with $h\ne 0$. It is interesting to note that
these solutions only exist for a certain range of deformations by ${\cal O}_h$ (at fixed $\mu$).
We then show that this one parameter family of
charged domain wall solutions arises as the zero temperature limit
of a more general class of superconducting black holes, once again generalising what was found in \cite{Gauntlett:2009dn}
for $h=0$. We first construct Reissner-Nordstr\"om-like charged black
holes\footnote{It is interesting to compare this class of charged black holes with
those that were very recently constructed in a top down model in \cite{Gubser:2009qt} and a phenomenological model in \cite{Goldstein:2009cv}.}
with $\chi=0$, but with $h\ne 0$, which
describe the high temperature normal phase of these superconductors.
At finite temperature these solutions have a regular horizon with
$h$ going to zero at the horizon. Furthermore, in the zero temperature limit,
for a certain range of deformations ${\cal O}_h\ne 0$, they approach $AdS_2\times \bbR^2$, exactly as
for the Reissner-Nordstr\"om black hole with $h=0$. For larger deformations, in the zero temperature limit they have vanishing entropy and become
singular.
We then show that for a certain range of deformations ${\cal O}_h\ne 0$ a new branch of black holes carrying charged scalar hair
appear and that they are thermodynamically favoured, thus demonstrating that we do indeed
have holographic superconductors. Our numerics indicate that the superconducting black holes exist for exactly the same class
of deformations where the Reissner-Nordstr\"om like black holes have an $AdS_2\times \bbR^2$ limit at zero temperature and hence 
at zero temperature the $AdS_2\times \bbR^2$ solutions are thermodynamically disfavoured.
Furthermore, we show that the solutions with charged hair smoothly map on the zero temperature
charged domain wall solutions with the Pope-Warner $AdS_4$ region in the IR, demonstrating the emergent $d=3$ conformal symmetry of these holographic superconductors.
It is worth highlighting that in the far IR all of these superconductors (for a given $SE_7$ space), when held at zero temperature and finite
chemical potential, are described by exactly the same universal CFT i.e. the CFT dual to the Pope-Warner $AdS_4$ vacuum.

If we assume that our analysis has captured all of the relevant instabilities in M-theory,
the phase diagram for our new holographic superconductors, for $\mu\ne 0$,  is summarised in Figure 1. The vertical axis is the temperature, the horizontal axis is the value of $h_1$ which determines the isotropic deformation by the operator ${\cal O}_h$. The most striking feature is the superconducting
dome that appears for $-h_1^c\le h_1 \le h_1^c$. Under this dome at zero temperature, and in the far IR, the superconductors
are all described by the same Pope-Warner $AdS_4$ solution. Outside of the dome at zero temperature the system is described
by singular solutions which require further investigation.  We will also construct black hole solutions with $\mu=0$ and this part of the phase diagram is incorporated in Figure 14 in the discussion section.
\begin{figure}[th]
\begin{center}
\includegraphics[width=0.6\textwidth]{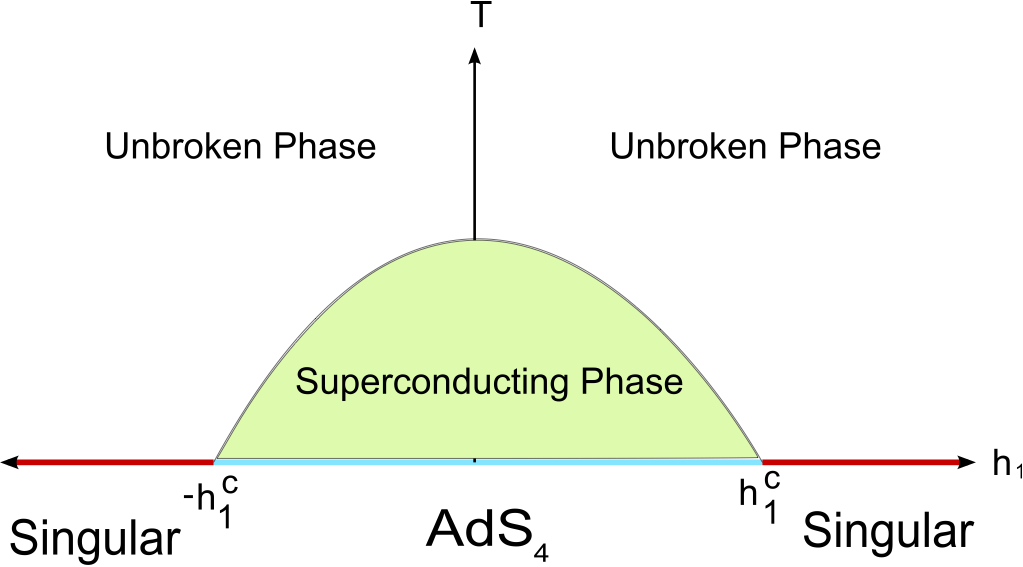}
\caption{The phase diagram for the holographic superconductors. The vertical axis is temperature, the horizontal axis determines the
deformation of the skew-whiffed CFT by the operator ${\cal O}_h$ and the chemical potential $\mu$ is non-zero.
\label{phasediag}}
\end{center}\end{figure}

Finally, we calculate the electrical conductivity of our black holes using linear response theory.
We find that the electrical conductivity contains both longitudinal and, when $h\ne 0$,
transverse (Hall) components for both the superconducting and normal phase black holes, the latter
arising from the broken parity and time reversal
invariance in the boundary theory.

The plan of the rest of the paper is as follows. In section 2 we briefly review the
consistent KK truncation on an arbitrary Sasaki-Einstein seven-manifold that was presented in \cite{Gauntlett:2009zw}.
In section 3 we discuss the three $AdS_4$ vacua of the $D=4$ theory and their uplifts to
$D=11$ solutions. Section 4 presents the new additional consistent KK truncation of the $D=4$ theory.
In section 5 we present the ansatz for the $D=4$ fields that we shall use to construct black hole and domain wall solutions.
We also discuss boundary counter terms in the action and some aspects of thermodynamics. As somewhat of an aside, section 6 discusses
uncharged domain walls corresponding to the RG flows between the $AdS_4$ vacua. Section 7 discusses the zero temperature limit solutions
of the charged black hole solutions that we construct in section 9. We construct the charged domain walls interpolating between
the deformed skew-whiffed $AdS_4$ vacuum and the Pope-Warner $AdS_4$ vacuum
and also the solutions interpolating between the skew-whiffed $AdS_4$ vacuum and the $AdS^2\times \bbR^2$.
Section 8 briefly discusses uncharged black hole solutions (i.e. $\mu= 0$) with $h\ne 0$. Section 9 discusses
both the normal and superconducting phase charged black hole solutions for general $h$
and presents some results on the conductivity of the black holes. We briefly conclude in section 10 and we have three appendices.

\section{The consistent KK truncation of \cite{Gauntlett:2009zw}  }

We start by summarising the consistent KK truncation of $D=11$ supergravity on
an arbitrary Sasaki-Einstein space $SE_7$ found in
\cite{Gauntlett:2009zw} (extending
\cite{Buchel:2006gb}\cite{Gauntlett:2007ma}).
First recall that any Sasaki-Einstein metric
can, locally, be written as a fibration over a six-dimensional K\"ahler-Einstein space
\be\label{semeth}
ds^2(SE_7)\equiv ds^2(KE_6)+\eta\otimes\eta
\ee
Here $\eta$ is the one-form dual to the Reeb Killing vector satisfying
$d\eta=2J$ where $J$ is the K\"ahler form of $KE_6$. We denote the $(3,0)$ form defined
on $KE_6$ by $\Omega$ and $d\Omega=4i\eta\wedge \Omega$.
The volume form is taken to be
$vol(SE_7)=\eta\wedge J^3/3!=(i/8)\eta\wedge\Omega\wedge\Omega^*$.

For a regular or quasi-regular Sasaki-Einstein manifold, the orbits of the
Reeb vector all close, corresponding to compact $U(1)$ isometry,
and the $KE_6$ is a globally defined manifold or
orbifold, respectively. For an irregular Sasaki-Einstein manifold, the Reeb-vector generates a
non-compact $\bbR$ isometry and the $KE_6$ is only locally defined.
For applications to holographic superconductivity one is most interested in cases with
$U(1)$ isometry.

In the KK ansatz the $D=11$ metric is written as
\be\label{KKmetT}
\frac{1}{(2L)^2}ds^2=e^{-6U-V}ds^2_4+e^{2U}ds^2(KE_6)+e^{2V}(\eta+A_1)\otimes(\eta +A_1) \ ,
\ee
while the four-form is written
\begin{equation}
\label{KKG4T}
\begin{aligned}
   \frac{1}{(2L)^3}G_4 &= 6 e^{-18U-3V}\left(
           \epsilon+h^2+|\chi|^2\right)\vol_4 + H_3 \wedge(\eta+A_1)
      + H_2 \wedge J
      \\ & \qquad
      + dh \wedge J \wedge (\eta+A_1) + 2h J \wedge J
      \\ & \qquad
      + {\sqrt 3}\left[
         \chi(\eta+A_1)\wedge\Omega-\tfrac{i}{4}D\chi\wedge\Omega
         + \textrm{c.c.} \right] \ .
\end{aligned}
\end{equation}
where $ds^2_4$ is a four-dimensional metric\footnote{Note that we are using the four-dimensional Einstein frame metric.},
$U,V,h$ are real scalars, $\chi$ is a
complex scalar defined on the four-dimensional space.
Furthermore, also defined on this four-dimensional space are $A_1$ a one-form potential, with field strength $F_2\equiv dA_1$,
two-form and three-form field strengths $H_2$ and $H_3$, related to one-form and two-form potentials
via
\bea\label{bianch}
H_3&=&dB_2\nn
H_2&=& dB_1+2B_2+h F_2
\eea
Note that the scalar $\chi$ is charged with respect to $A_1$ and in particular we have
$D\chi\equiv d\chi-4iA_1\chi$.
Also note that $\epsilon$ appearing in the four-form flux is a constant:
\be
\epsilon=\pm1
\ee
whose significance will be explained below.
Our conventions for $D=11$ supergravity are as in
\cite{gp}; in particular we note that the $D=11$ volume form is
given by $vol_4\wedge vol(SE_7)$.
Finally, $L$ is an arbitrary length scale which we will later set to 1/2.

This provides a consistent KK truncation of $D=11$ supergravity in the sense that if one finds a solution
to the equations of motion for the $D=4$ metric $g$ and matter fields $U,V,A_1,B_1,B_2,h,\chi$,
as given in \cite{Gauntlett:2009zw}, then one has found a solution to the $D=11$ supergravity equations
of motion.
The $D=4$ equations of motion can be derived from an action given in \cite{Gauntlett:2009zw}.
In this paper we will find it convenient to work with an action that is obtained after dualising
the one-form $B_1$ to another one form $\tilde B_1$ and the two-form $B_2$ to a scalar $a$
as explained in section 2.3 of \cite{Gauntlett:2009zw}. The dual action is given by
\begin{equation}
\label{lageinfull}
\begin{aligned}
  S &= \frac{(2L)^2}{16\pi G}\int d^4 x\sqrt{-g}\Big[
      R - 24(\nabla U)^2 -\tfrac32(\nabla V)^2 - 6\nabla U\cdot\nabla V
      - \tfrac{3}{2}e^{-4U-2V}(\nabla h)^2
        \\&\qquad\quad
        - \tfrac32 e^{-6U} |D\chi|^2
 - \tfrac{1}{4}e^{6U+3V}F_{\mu\nu}F^{\mu\nu}
      -\tfrac{3}{4}\frac{e^{2U+V}}{4h^2+e^{4U+2V}}(\tilde H_2+h^2F_2)_{\mu\nu}(\tilde H_2+h^2F_2)^{\mu\nu}
      \\&\qquad\quad
      -\tfrac{1}{2}e^{-12U}\left[\nabla a+6(\tilde B_1-\epsilon A_1)-\tfrac{3}{4}i(\chi^* D\chi-\chi D\chi^*)\right]^2
      + 48e^{-8U-V} - 6e^{-10U+V}
      \\ & \qquad \quad
      - 24h^2e^{-14U-V}
      - 18\left( \epsilon+h^2 +|\chi|^2 \right)^2e^{-18U-3V}
      - 24e^{-12U-3V} |\chi|^2
      \Big] \\ &
      +\frac{(2L)^2}{16\pi G}\int \Big[ - h^3 F_2\wedge F_2 -3h \tilde H_2\wedge F_2 +\frac{3h}{(4h^2+e^{4U+2V})}
      (\tilde H_2+h^2F_2)\wedge(\tilde H_2+h^2F_2)
      \Big] \ .
\end{aligned}
\end{equation}
where $\tilde H_2\equiv d\tilde B_1$. The dual fields are related to the original fields via
\bea
H_2&=&\frac{1}{4h^2+e^{4U+2V}}\left[2h(\tilde H_2+h^2 F_2)-e^{2U+V}*(\tilde H_2+h^2 F_2)\right]\nn
H_3&=&-e^{-12U}*\left[da+6(\tilde B_1-\epsilon A_1)+\tfrac{3}{4}i(\chi^* D\chi-\chi D\chi^*)\right]
\eea
Note that the factors of $L$ appear in a more conventional way if one
uses the rescaled metric $\tilde g=(2L)^2 g$.

\section{$AdS_4$ vacuum solutions}
The $D=4$ equations of motion arising from \reef{lageinfull}
admit various $AdS_4$ solutions with $F_2=\tilde H_2=a=0$ and
for various values of the scalar fields $U,V,h$ and $\chi$.
In the following $ds^2(AdS_4)$ will always denote the standard unit radius
$AdS_4$ metric and $Vol(AdS_4)$ the corresponding volume-form as given in appendix A of
\cite{Gauntlett:2009zw}.
We will determine the masses of the other fields considered as perturbations
around each $AdS_4$ solution to obtain the scaling dimensions of the dual operators
in the boundary CFT. For scalar fields with mass $m$ the scaling dimensions are given by
\be\label{ss}
\Delta=\frac{3}{2}\pm \frac{1}{2}[9+4m^2 R^2_{AdS}]^{1/2}
\ee
while those for vector fields are given by
\be\label{sv}
\Delta=\frac{3}{2}\pm \frac{1}{2}[1+4m^2 R^2_{AdS}]^{1/2}
\ee
where $R^2_{AdS}$ is the radius squared of the Einstein frame $AdS$ metric $g$.
Note that if we used the metric $\tilde g= (2L)^2g$ then $R^2_{AdS}\to (2L)^2R^2_{AdS}$ and
$m^2\to m^2/(2L)^2$.

\subsection{$AdS_4\times SE_7$ solutions: supersymmetric and skew-whiffed}
The simplest $AdS_4$ vacua have $\epsilon=\pm 1$,
\bea
U=0,\quad {V}=0, \quad \chi=0,\quad h=0
\eea
and the radius squared of the Einstein $AdS_4$ metric is given by
\be
R^2_{AdS}=\frac{1}{4}.
\ee
These uplift to the $D=11$ solutions:
\bea\label{d11vac}
\frac{1}{(2L)^2}ds^2&=&\tfrac{1}{4}ds^2(AdS_4)+ds^2(SE_7)\nn
\frac{1}{(2L)^3}G_4&=&\epsilon\tfrac{3}{8}Vol(AdS_4)
\eea
When $\epsilon=+1$, these $AdS_4\times SE_7$ solutions are supersymmetric and are dual to $d=3$ SCFTs with,
generically, ${\cal N}=2$ supersymmetry. 
For these solutions the Killing vector dual to the one-form $\eta$ in
the $SE_7$ metric \reef{semeth} is dual to an $R$-symmetry.
On the other hand when $\epsilon=-1$ the solutions
are skew-whiffed $AdS_4\times SE_7$ solutions and generically do not preserve any supersymmetry at all.
An important exception is when $SE_7$ is the round seven-sphere in which case both $AdS_4$ solutions
preserve maximal supersymmetry. The skew-whiffed solutions have been shown to be perturbatively stable
\cite{Duff:1984sv} and thus should be dual to well defined CFTs at least in the supergravity approximation. Some discussion on
the possibility of $1/N$ effects destabilising the non-supersymmetric skew-whiffed
solutions has been discussed in \cite{Berkooz:1998qp}\cite{am} and it would be interesting
to explore this further. 
Note that for the skew-whiffed $AdS_4\times SE_7$ 
solutions the Killing vector dual to the one-form $\eta$ in
the $SE_7$ metric \reef{semeth} is dual to a global symmetry in the dual CFT
and is an $R$-symmetry just for the case of $SE_7=S^7$.

The spectrum of the $D=4$ theory in these backgrounds was discussed in \cite{Gauntlett:2009zw}.
For $\epsilon=+1$, $m^2_h=40$, $m^2_\chi=40$ and $U,V$ mix to give $m^2=16,72$. These
give scaling dimensions $\Delta=5,5$ and $\Delta=4,6$ respectively. There is also a massless gauge field
and a massive gauge field with $m^2=48$ corresponding to $\Delta=2$ and $\Delta=5$, respectively.

When $\epsilon=-1$, $U,V$ mix to again give $m^2=16,72$, corresponding to scaling dimensions
$\Delta=4,6$ respectively. On the other hand now $m^2_h=m^2_\chi=-8$ with each corresponding to scaling dimension
$\Delta_\pm=1,2$. The masses of the gauge fields are unchanged.
Note that in both cases the field $a$ becomes the longitudinal mode of the massive gauge field.

\subsection{Pope-Warner solutions}
Another $AdS_4$ vacuum is obtained when $\epsilon=-1$,
\bea
e^U=2^{-1/6},\quad e^{V}=2^{1/3}, \quad \chi^2=2/3,\quad h=0
\eea
and the radius squared of the Einstein $AdS_4$ metric is given by
\be
R^2_{AdS}=\frac{3}{16}
\ee
Choosing $\chi=+(2/3)^{1/2}$ for definiteness we find that this uplifts to the $D=11$ class solutions
\bea
\frac{1}{(2L)^2}ds^2&=&2^{2/3}\left[\tfrac{3}{16}ds^2(AdS_4)+\tfrac{1}{2}ds^2(KE_6)+\eta\otimes\eta\right]\nn
\frac{1}{(2L)^3}G_4&=&2\left[-\tfrac{9}{64}Vol(AdS_4)+\tfrac{1}{\sqrt 2}(\eta\wedge\Omega+c.c)\right]
\eea
which were first constructed in \cite{Pope:1984bd}\cite{Pope:1984jj}.
It has been shown that they do not preserve any supersymmetry but a stability analysis has not yet been performed.
Note that these solutions are topologically $AdS_4\times SE_7$: the fibre of the $SE_7$ metric
being stretched by a factor of $\sqrt 2$ compared to the $SE_7$ metric in \reef{semeth}.

In this background, the $D=4$ theory gives two massive scalars with
$m^2=32$ and two with $m^2=96$ corresponding to scaling dimensions
$\Delta=3/2+{\sqrt 33}/2$ and $\Delta=6$, respectively.
There are also two massive gauge fields with $m^2=32$ and $m^2=96$ corresponding to scaling dimensions
$\Delta=4$ and $\Delta=3/2+{\sqrt 73}/2$, respectively. Note that the phase of $\chi$ and the
field $a$ become longitudinal modes of the massive gauge-fields.

\subsection{Englert solutions}
Another $AdS_4$ vacuum is obtained when $\epsilon=-1$,
\bea
e^U=(4/5)^{1/6},\quad e^{V}=(4/5)^{1/6}, \quad \chi^2=4/15,\quad h^2=1/5
\eea
and the radius squared of the Einstein $AdS_4$ metric is given by
\be
R^2_{AdS}=\frac{12}{25{\sqrt 5}}
\ee
Choosing $\chi=+(2/15)^{1/2}$, $h=+1/{\sqrt 5}$ for definiteness we find that this uplifts to the
$D=11$ class of solutions
\bea
\frac{1}{(2L)^2}ds^2&=&(\tfrac{4}{5})^{1/3}\left[\tfrac{3}{10}ds^2(AdS_4)+ds^2(KE_6)+\eta\otimes\eta\right]\nn
\frac{1}{(2L)^3}G_4&=&(\tfrac{4}{5})^{1/2}
\left[-\tfrac{9}{25}Vol(AdS_4)+J\wedge J+(\eta\wedge\Omega+c.c)\right]
\eea
This an Englert-type solution and note that the metric on $SE_7$ is exactly the same as
in \reef{semeth}.
For the special case when $SE_7=S^7$ it was first constructed in
\cite{Englert:1982vs} and the generalisation was suggested in \cite{Awada:1982pk}.
This solution is known not to preserve any supersymmetry (for the
$S^7$ case this was shown in \cite{Englert:1983qe} and the results of \cite{Page:1984fu} show this more generally) and to
be unstable \cite{Page:1984fu}.

In this background, the $D=4$ theory gives four massive scalars with
$m^2=-5{\sqrt 5}$, $25{\sqrt 5}/2$, $20{\sqrt 5}$, $75{\sqrt 5}/2$. Note that the first mode has complex scaling dimension and thus
violates the BF bound. This unstable mode, as well as the other three modes, are precisely the same modes
considered in \cite{Page:1984fu}. The scaling dimensions of the three other scalars are $\Delta=3/2+{\sqrt 33}/2, 3/2+(237/20)^{1/2}, 6$, respectively.
There are also two massive gauge fields with $m^2=5{\sqrt 5}$ and $m^2=30{\sqrt 5}$ corresponding to scaling dimensions
$\Delta=3/2+(53/20)^{1/2}$ and $\Delta=3/2+(293/20)^{1/2}$, respectively.

\subsection{Flux quantisation and central charges}
The $D=11$ equation of motion for the four-form is $d*G_4+\tfrac{1}{2}G_4\wedge G_4=0$ and the quantised membrane charge
is given by
\be
N=\frac{1}{(2\pi l)^6}\int(*G_4+\tfrac{1}{2} C_3\wedge G_4)
\ee
where $dC_3=G_4$ and $l$ is the $D=11$ Planck length. For each of the above solutions with $\epsilon=-1$, we find
\bea\label{enns}
N&=&6\left(\frac{(2L)}{2\pi l}\right)^6vol(SE_7)
\eea
Note that in our conventions this is counting the number of anti-membranes.
We can also calculate the central charge using the formula \cite{Kovtun:2008kw}
\be
c\equiv 384\frac{(2L)^2R_{AdS}^2}{16\pi G}
\ee
In our conventions the D=11 action is
\bea
S&=&\frac{1}{(2\pi)^8l^9}\int d^{11}x\sqrt{-g_{11}}[R_{11}+...]
\eea
and hence
\be
\frac{1}{16\pi G}=\frac{(2L)^7vol(SE_7)}{(2\pi)^8l^9}
\ee
We thus deduce
\bea
c&=&\frac{128\pi N^{3/2}}{6^{1/2}vol(SE_7)^{1/2}}R^2_{AdS}
\eea

By comparing $R^2_{AdS}$ of the different $AdS_4$ vacua, one concludes that it might be possible to
find a domain wall solution that interpolates between a perturbed skew-whiffed vacuum
in the UV and the Pope-Warner vacuum in the IR, which would be dual to an RG flow between the corresponding dual CFTs. We shall see later that this
is indeed the case. In fact we will see that there is a one parameter family
of domain walls that interpolates between these vacua. We will also find
a domain wall solution that interpolates from the skew-whiffed vacuum in the UV to the Englert vacuum
in the IR and another domain wall solution that interpolates between the Englert vacuum in the UV and
the Pope-Warner in the IR. The instability of the Englert vacuum makes these
latter solutions less interesting as far as RG flows are concerned.

\section{Further consistent KK truncation}
When $\epsilon=-1$ there is a further consistent truncation of the $D=4$
theory described by \reef{lageinfull} that is obtained by setting
\bea\label{fac}
a&=&0\nn
\tilde B_1&=&-A_1\nn
e^{6U}&=&1-\tfrac{3}{4}|\chi|^2\nn
e^{6V}&=&\frac{(1-h^2)^3}{(1-\tfrac{3}{4}|\chi|^2)^2}
\eea
Note that this implies
\bea
H_3&=&\frac{3i}{4(1-\tfrac{3}{4}|\chi|^2)^2}*[\chi^*D\chi-\chi D\chi^*]\nn
H_2&=&\frac{(1-h^2)}{1+3h^2}[-2hF_2+(1-h^2)^{1/2}*F_2]
\eea
After substituting this into the equations of motion derived from
\reef{lageinfull} (see appendix B of \cite{Gauntlett:2009zw})
we obtain equations that can be derived from the following action
\bea
\label{lagred}
  &&S = \frac{1}{16\pi G}\int d^4 x\sqrt{-g}\Big[
      R - \frac{(1-h^2)^{3/2}}{1+3h^2}{F}_{\mu\nu}
{F}^{\mu\nu}
-\frac{3}{2(1-\tfrac{3}{4}|\chi|^2)^2}|D\chi|^2\nn
&&-\frac{3}{2(1-h^2)^2}(\nabla h)^2
-\frac{24(-1+h^2+|\chi|^2)}{(1-\tfrac{3}{4}|\chi|^2)^2(1-h^2)^{3/2}}
      \Big]+\frac{1}{16\pi G}\int \frac{2h(3+h^2)}{(1+3h^2)}F\wedge F\nn
\eea
This action is also obtained by substituting the ansatz \reef{fac} directly
into \reef{lageinfull}. From now on we have set $L=1/2$. Observe that
we must have $|\chi|<\tfrac{2}{\sqrt{3}}$ and $|h|<1$.
We also observe that all terms in the Lagrangian
except for the coupling of $h$ to $F\wedge F$ are invariant under
$h\to -h$. As we shall see later this coupling
leads to breaking of parity and time reversal invariance in the dual CFT.

It is worth pointing out that there are further consistent truncations that one can consider.
For example, one can consistently set $\chi=0$ to obtain a theory of gravity coupled to
a gauge field $A_1$ and a neutral scalar $h$, and we will find black hole solutions of this theory
(it is interesting to compare and contrast with the phenomenological models studied in \cite{Gubser:2009qt}\cite{Goldstein:2009cv}).
It is also possible to then further consistently set $A_1=0$. Alternatively, we can set $A_1=0$ provided that we
restrict $\chi$ to be real: $\chi=\chi_R$.
This theory with two real scalars can then be
further consistently truncated\footnote{Observe that this latter
theory of a single real scalar field can be obtained
from the truncation considered in equation (3.3) of
\cite{Gauntlett:2009zw}. In particular, we obtain the same theory
after setting $e^{2v}=1-h^2$ in (3.3) of \cite{Gauntlett:2009zw}, after
flipping the sign of 3 in the last term in the potential which is needed
for the skew-whiffed case with $\epsilon=-1$.}
by either setting $\chi_R=(2/{\sqrt 3})h$ or by setting $h=0$.
Finally, observe that we can set $h=0$ provided that  we impose $F\wedge F=0$ by hand \cite{Gauntlett:2009dn} (and hence setting $h=0$ is not a consistent KK truncation).

Observe that all of the $AdS_4$ vacua discussed in the last subsection are
solutions of the consistent truncation \reef{lagred}.
For the skew-whiffed $AdS_4$ solution
the perturbed fields $\delta\chi$ and $\delta h$ have masses given by
$m^2_\chi=m^2_h=-8$, as before, and the gauge field is massless.
For the Pope-Warner $AdS_4$ solution we have $m^2_\chi=m^2_h=32$ and the
gauge field is massive with $m^2=32$. For the Englert solution
we find that the scalar perturbations mix and
that $\delta h-({\sqrt 3}/2)\delta \chi$ and
$\delta \chi+({\sqrt 3}/2)\delta h$ have mass squared $-5{\sqrt 5}$ and
$25{\sqrt 5}/2$, respectively.
In particular, the unstable, BF violating mode about the Englert solution
is contained within the truncation \reef{lagred}. The gauge field is massive
in the Englert solution with $m^2=5{\sqrt 5}$.

In the skew-whiffed vacuum $h,\chi$ are dual to operators ${\cal O}_h$, ${\cal O}_\chi$
each with conformal dimension $\Delta_\pm =1,2$.
It is worth emphasising that in the solutions of the truncated theory \reef{lagred}
with $h,\chi\ne 0$, in addition to activating these operators in the dual CFT we are also
activating the operators with $\Delta=4,6$ that are dual to linear combinations of the fields $U,V$,
via \reef{fac}. More precisely, in the solutions we shall consider that asymptotically approach the skew-whiffed $AdS_4$ vacuum, the asymptotic falloffs of $h,\chi$ will include cases where the skew-whiffed CFT is deformed by ${\cal O}_h$ and,
for the RG flows in section 6 only, by ${\cal O}_\chi$, and also where these operators acquire vevs.
This means that the operators dual to $U,V$ are not deforming the skew-whiffed CFT but they are
acquiring vevs which can easily be worked out for our solutions. However, we will not include the details.

Finally, we observe that if we  set $h=0$ in the truncated action \reef{lagred}
(which is only possible for configurations with $F\wedge F=0$), and then linearize in
$|\chi|$, we make contact with the model with a simple mass term considered in \cite{H32}
(after rescaling their gauge field by a factor of 2 and setting their $q=2$. Also when
$h=0$ the action \reef{lagred} is in the class considered in \cite{Franco:2009yz}.

\section{Ansatz for black hole and domain wall solutions}
For the remainder of the paper we will consider the following ansatz for the $D=4$ fields in
the truncated theory described by the action \reef{lagred}.
For the metric we take
\newcommand\R{{r}}
\newcommand\G{{g}}
\newcommand\B{{\beta}}
\be\label{mans}
ds^2=-\G e^{-\B}dt^2+\G^{-1}d\R^2+\R^2(dx^2+dy^2)
\ee
where $g,\beta$ are functions of $r$ only.
The gauge-field $A_1$ is taken to be purely electric
\bea\label{gans}
A_1&=&\phi(r) dt\ ,
\eea
and we will also impose
\be\label{sans}
\chi\equiv\xi(r)\in \bbR,\qquad h=h(r)
\ee

We now substitute into the equations of motion arising from the action
\reef{lagred}. After some calculation we are led
to five ordinary differential equations, which we have presented in
\reef{phieqe},
\reef{rhoeqe},
\reef{heqe},
\reef{eone} and \reef{etwo}, for five real functions $\phi,\xi,h,g$ and $\beta$.
These equations can also be obtained from an action obtained by substituting the above
ansatz directly in to the action \reef{lagred}:
\bea\label{actansatz}
S&=&c_0\int dr r^2e^{-\beta/2}\Big[-g''+g'(\frac{3}{2}\beta'-\frac{4}{r})+g(\beta''-\frac{1}{2}(\beta')^2+2\frac{\beta'}{r}-\frac{2}{r^2})\nn
&&
\qquad\qquad-\frac{3(g\xi'^2-16g^{-1}
e^{\beta}\phi^2\xi^2)}{2(1-\tfrac{3}{4}\xi^2)^2}
-\frac{3g(h')^2}{2(1-h^2)^2}
+\frac{2(1-h^2)^{3/2}e^{\beta}\phi'^2}{1+3h^2} \nn
&&\qquad\qquad-\frac{24(-1+h^2+\xi^2)}{(1-\tfrac{3}{4}\xi^2)^2 (1-h^2)^{3/2}}\Big]
\eea
where
\be
c_0\equiv\frac{1}{16\pi G}\int dt dx dy
\ee
It will be helpful to note the following scaling symmetries:
\be
r\to ar,\quad(t,x,y)\to a^{-1}(t,x,y),\quad g\to a^2g,\quad
\phi\to a\phi,
\ee
and
\be
e^\beta\to a^2e^{\beta},\quad t\to at,\quad \phi\to a^{-1}\phi
\ee
which leave the metric, $A_1$, and all equations of motion
invariant. Notice that this ansatz has the symmetry $h\to -h$. We also have
the $\bbZ_2$ symmetries $\xi\to -\xi$
and $\phi\to-\phi$. Also notice that it is consistent to separately set $\xi=0$, $\phi=0$ or $h=0$.

We will be almost exclusively interested in solutions that asymptote to a perturbation of the skew-whiffed
$AdS_4$ vacuum. We recall that a scalar field dual to an operator in the CFT with scaling dimension $\Delta$
has the two asymptotic behaviours
\be
r^{\Delta -3}\qquad {\rm and}\qquad r^{-\Delta}
\ee
whereas a vector field behaves as
\be
r^{\Delta - 2}\qquad {\rm and}\qquad r^{1-\Delta}
\ee
We thus focus on the asymptotic expansion
\footnote{To compare with \cite{Gauntlett:2009dn} we should
identify $\varepsilon,\mu,q,\xi_i$ with $m,\hat\mu,\hat q,\sigma_i$, respectively.  }
\bea\label{aesw}
g&=& 4 r^2 +16 \pi G \left({{h_1}}^2+{{\xi_1}}^2\right)
- 8\pi G[\varepsilon-4\xi_1 \xi_2 -4 h_1 h_2]\frac{1}{r}+\dots\nn
\beta & = & {{\beta_a}}+4\pi G \left({{h_1}}^2+{{\xi_1}}^2\right) \frac{1}{r^2}
+\frac{32\pi G}{3} ({{h_1}} {{h_2}}+{{\xi_1}} {{\xi_2}}) \frac{1}{r^3}+\dots\nn
\xi & = & {\sqrt \frac{16\pi G}{3}}\left[\frac{\xi_1}{r}+\frac{\xi_2}{r^2}+\dots\right]\nn
h & = & {\sqrt \frac{16\pi G}{3} }\left[\frac{h_1}{r}+ \frac{h_2}{r^2}+\dots\right]\nn
\phi & = &{\sqrt {4\pi G}}e^{-\beta_a/2}\left[ \mu -\frac{q}{r}+\dots\right]
\eea
It is worthwhile noting that
\begin{eqnarray}
e^{-\beta} g= e^{-\beta_a}\left[4r^2- \frac{8\pi G[\varepsilon+\tfrac{4}{3}\xi_1\xi_2+\tfrac{4}{3}h_1h_2]}{r}
+\dots\right]
\end{eqnarray}

\subsection{Skew-whiffed to Pope-Warner uncharged and charged domain walls}
As we will discuss in more detail later, we will be interested in both charged and
uncharged domain wall solutions that asymptote in the IR to the Pope-Warner vacuum as $r\to 0$.
Hence we will consider the following expansion
\bea\label{irexp}
\xi    &=& \sqrt{\frac{2}{3}} + a_\xi r^{\Delta_{\rm PW}^\xi -3}+ \cdots\nonumber \\
h   &=& a_h r^{\Delta_{\rm PW}^h-2} + \cdots  \nonumber\\
\beta &=& a_\beta+ \cdots \nonumber\\
g &=& \frac{16 r^2}{3}    + \cdots\nonumber \\
\phi &=& a_\phi r^{\Delta_{\rm PW}^\phi-2} + \cdots
\eea
where $\Delta_{\rm PW}^\xi=\Delta^h_{PW}=
(3+{\sqrt 33})/2$ and $\Delta_{\rm PW}^\phi=4$.
Here $\{ a_\xi, a_h , a_\beta, a_\phi \}$ parametrises
all possible marginal and irrelevant operators of the Pope-Warner vacuum within our truncation.

\subsection{Skew-whiffed to $AdS_2 \times \mathbb{R}^2$ solutions}
The Reissner-Nordstr\"om electrically charged black hole solution (with planar horizon) is given by,
\be\label{rnsol}
g=4r^2-\frac{1}{r}(4r_+^3+\frac{\alpha^2}{r_+})+\frac{\alpha^2}{r^2},\qquad \phi=\alpha(\frac{1}{r_+}-\frac{1}{r})
\ee
with $h = \xi = \beta = 0$, where $\alpha=\sqrt{4\pi G} q$, $r_+=q/\mu$.
At zero temperature, when $\alpha^2=12r_+^4$ and $g=4(r-r_+)^2(r^2+2r_+r+3r_+^2)/r^2$,
the near horizon geometry of this solution is given by $AdS_2\times \bbR^2$ with the $AdS_2$ having radius squared $1/{24}$.
In other words, at zero temperature this charged solution interpolates between skew-whiffed $AdS_4$ in the UV and
$AdS_2\times \bbR^2$ in the IR.

Later we will consider deformations of this interpolating solution by marginal and irrelevant operators with respect to the $AdS_2$ factor in the IR
which allow us to find zero temperature charged solutions with $h\ne 0$ and $\chi=0$
which again interpolate from a deformed skew-whiffed $AdS_4$ vacuum in the UV to an $AdS_2\times \bbR^2$ geometry in the IR.
Specifically we find the $AdS_2\times \bbR^2$ IR geometry is
asymptotically given by
\bea\label{irexp2}
h   &=& h_+ (r - r_+)^{(\sqrt{105} - 3)/6} + \cdots  \nonumber\\
\phi &=& 2 \sqrt{3} e^{-\beta_+/2} \left( \frac{1}{r_+} - \frac{1}{r}
\right) + \cdots  \nonumber\\
g &=& 4 r^2 - \frac{1}{r} \left( 4 r_+^3 + \frac{12}{r_+} \right) +
\frac{12}{r^2} + \cdots \nonumber \\
\beta &=& \beta_+ + \cdots
\eea
where $\beta_+$ parameterizes the only marginal deformation, and $h_+$
the only irrelevant deformation in our ansatz.

\subsection{Black hole solutions}
For the general black hole solutions we demand that there is a regular finite temperature horizon located at $r=r_+$.
Specifically we impose that
\bea
g(r_+)=\phi(r_+)=0
\eea
and so the solution is specified by five parameters at the horizon
\be\label{horpar}
r_+,\quad \beta_+=\beta(r_+),\quad \phi_+=\phi'(r_+),\quad\xi_+= \xi(r_+),\quad h_+=h(r_+)
\ee
(we will return to this counting later).
We have the expansion as $r\to r_+$
\bea\label{exphor}
\xi &=& \xi_+ + \xi_+^{(1)}(r-r_+) + \cdots \nonumber\\
h &=& h_+ + h_+^{(1)}(r-r_+) + \cdots \nonumber\\
\phi &=& \phi_+(r-r_+) + \phi_+^{(2)}(r-r_+)^2\cdots\nonumber\\
g &=&g_+^{(1)}(r-r_+) + g_+^{(2)}(r-r_+)^2\cdots\nonumber\\
\beta &=& \beta_+ + \beta_+^{(1)}(r-r_+) + \cdots
\eea
where for example,
\bea
g_+^{(1)} = \frac{r_+\left[ 12(1 + 3 h_+^2)(-1 + h_+^2 +\xi_+^2) + e^{\beta_+}\phi_+^2(1-h_+^2)^3 \left(1-\tfrac{3}{4}\xi_+^2\right)^2\right]}{\sqrt{1-h_+^2}\left( 1 - \frac{3}{4}\xi_+^2 \right)^2 \left( -1 - 2h_+^2 + 3 h_+^4 \right)}
\eea
and analogous expressions can be obtained for
$g_+^{(2)},\xi_+^{(1)},h_+^{(1)},\beta_+^{(1)},\phi_+^{(2)},\dots$ in terms of the data given in \reef{horpar}.

\subsection{Counter terms and black hole thermodynamics}
To calculate thermodynamic quantities for the black hole solutions
we would like to calculate the on-shell Euclidean action $I$. As usual this will require
adding counter terms which are also relevant for the domain wall solutions.
Interesting early work analysing the thermodynamics of $R$-charged Reissner-Nordstr\"om black holes
using AdS/CFT techniques was carried out in \cite{Chamblin:1999tk}\cite{Cvetic:1999ne}.

We analytically continue by writing
\be
I=-iS,\qquad t=-i\tau
\ee
The temperature of the black hole is $T=e^{\beta_a/2}/\Delta \tau$
where $\Delta \tau$ is fixed by demanding regularity of the Euclidean
metric at $r=r_+$. We find using \reef{eone} and \reef{exphor} that
\bea
T&=&\frac{e^{\beta_a/2}}{4\pi}\left[g'e^{-\beta/2}\right]_{r=r_+}\nn
&=&\frac{r_+e^{\beta_a/2}}{4\pi}\left[
\frac{12e^{-\beta/2}(1-h^2-\xi^2)}{(1-\tfrac{3}{4}\xi^2)^2 (1-h^2)^{3/2}}
-\frac{e^{\beta/2}(1-h^2)^{3/2}}{1+3h^2}(\phi')^2\right]_{r=r_+}
\eea
As explained in appendix B, we find that the on-shell action can be expressed as
\bea\label{actone}
I_{OS}&=&\frac{\Delta \tau vol_2}{16\pi G}\int_{r_+}^{\infty}d\R \left[2\R\G e^{-\B/2}\right]'
\eea
where $\vol_2\equiv \int dx dy$.
Since $g(r_+)=0$, this expression only gets contributions from the on-shell
functions at $r=\infty$. An alternative expression for the on-shell action is given by
\bea\label{acttwo}
I_{OS}&=&\frac{\Delta \tau vol_2}{16\pi G}\int_{r_+}^{\infty}d\R
\left[r^2e^{-\beta/2}(g'-g\beta'- 4e^\beta\frac{(1-h^2)^{3/2}}{1+3h^2}\phi\phi'
)\right]'
\eea
which gets contributions from both $r=r_+$ and $r=\infty$.

\newcommand\ce{{\frac{\Delta \tau vol_2}{16\pi G}}}

The on-shell action diverges and we need to regulate by adding appropriate counter terms.
By examining the asymptotic expansion of the fields given in \reef{aesw}, we find
that the following counter-term action renders the total action finite:
\be
I_{ct}=\frac{1}{16\pi G}\int d\tau d^2x {\sqrt{g_\infty}}\left[-2K+8+3\xi^2+3h^2\right]
\ee
where $\sqrt{g_\infty}=\lim_{r\to\infty}g^{1/2}r^2e^{-\beta/2}$
and $K=\lim_{r\to\infty}g^{\mu\nu}\nabla_\mu n_\nu$
is the trace of the extrinsic curvature.
For the class of solutions under consideration we find
\be\label{ctermact}
I_{ct}=\frac{\Delta\tau vol_2}{16\pi G}\lim_{r\to\infty}e^{-\beta/2}\left[-r^2e^{\beta}(ge^{-\beta})'-4rg+r^2g^{1/2}(8+3\xi^2+3h^2)\right]
\ee
Defining
\be\label{itotal}
I_{Tot}=I+I_{ct}
\ee
we find that corresponding to the two expressions \reef{actone}, \reef{acttwo},
the on-shell total action can be written
as
\bea\label{onshellact}
[I_{Tot}]_{OS}
&=&\frac{vol_2}{T}\left(-\tfrac{1}{2}{\varepsilon}-{2}{\xi_1 \xi_2}
-{2}{h_1 h_2}\right)\nn
&=&\frac{vol_2}{T}\left(
\varepsilon-\mu q-Ts\right),
\eea
respectively,
where we have defined the entropy density $s$ to be
\be
s=\frac{(r_+^2)}{4G}
\ee
(the total entropy is $s vol_2 $).
The equality of these two expressions imply the Smarr-type relation
\be\label{smarr}
\tfrac{3}{2}\varepsilon=\mu q +Ts -{2}\xi_1\xi_2 -{2}h_1h_2
\ee

A variation of the action $I$
yields the equations of motion together with surface terms. For an
on-shell variation the only terms remaining are the surface terms
\begin{eqnarray}
\delta I _{OS}  & = & \frac{\Delta \tau vol_2}{16 \pi G}
\int dr  \partial_r \Big\{ \delta g \left[ e^{-\beta/2} r ( 2- r \beta' )\right]
+  \delta g' \left[ e^{-\beta/2} r^2 \right]
\nonumber \\
&& +  \delta \beta \left[  \tfrac{1}{2} e^{-\beta/2} r^2 ( g \beta' -g'  ) \right]
-  \delta \beta'  \left[ e^{-\beta/2} r^2 g \right]
\nonumber \\
&& + \delta \xi \left[ \frac{3 r^2e^{-\beta/2}g}{(1-\tfrac{3}{4}\xi^2)^2 }\xi'  \right]
+ \delta h \left[ 3 r^2e^{-\beta/2}g\frac{h'}{(1-h^2)^2}  \right]
\nn
&& - \delta \phi \left[ 4r^2e^{\beta/2} \frac{(1-h^2)^{3/2}}{1+3h^2}
\phi' \right]
 \Big\}
\end{eqnarray}
In the Euclidean black hole solution the only boundary is the conformal boundary $r \rightarrow \infty$ and  hence this integral only gets contributions there.
In addition one must also add the variation of the counter terms,
\begin{eqnarray}
\delta I_{ct} & = & \frac{\Delta\tau \, vol_2}{16 \pi G}
\lim_{r \rightarrow \infty}
\Big\{ \delta g \left[ e^{-\beta/2} r    (  r \beta' - 4 + \tfrac{1}{2} r (8+3 \xi^2+3h^2) g^{-1/2}  )  \right] -  \delta g' \left[ e^{-\beta/2} r^2 \right]
\nonumber \\
&& +  \delta \beta \left[  \tfrac{1}{2}re^{-\beta/2}  (  r g' -  rg \beta' +4g
-  r g^{1/2} (8 + 3 \xi^2 +3h^2)) \right]
+  \delta \beta'  \left[ e^{-\beta/2} r^2 g \right]
\nonumber \\
&& + \delta \xi \left[  6 e^{-\beta/2} r^2 g^{1/2} \xi\right]  + \delta h \left[  6 e^{-\beta/2} r^2 g^{1/2} h
\right]
 \Big\}
\end{eqnarray}
Combining these expressions we deduce that
\bea\label{vp}
[\delta I_{Tot}]_{OS}={\Delta\tau vol_2}e^{-\beta_a/2}
\left[(-\tfrac{1}{2}\varepsilon +\tfrac{1}{2}\mu q)\delta\beta_a
-q\delta \mu-{4}\xi_2\delta\xi_1-{4}h_2\delta h_1\right]
\eea
(which corrects equation (17) of \cite{Gauntlett:2009dn} by a factor of $16\pi$).
Note that we are keeping $\Delta\tau$ fixed in this variation, and hence $\delta \beta_a = 2 \, \delta T / T$.
Hence we see that $I_{Tot}$
is stationary for fixed temperature and chemical potential (i.e. $\delta \beta_a = \delta \mu = 0$) and for either $\xi_2 = 0$ or fixed $\xi_1$
and similarly either $h_2=0$ or fixed $h_1$. In our applications we will always fix
$\xi_1$ and $h_1$.

We now define the thermodynamic potential for a grand canonical ensemble via
$W\equiv T[I_{Tot}]_{OS}\equiv w vol_2$. From the above variations we see
that $w=w(T,\mu,\xi_1,h_1)$
and using the second expression in \reef{onshellact} we deduce the first
law
\be
\delta w=-s\delta T-q\delta\mu-4\xi_2\delta \xi_1 -4h_2\delta h_1
\ee
Note that from the second expression in \reef{onshellact} we can
write $w=\varepsilon-Ts-\mu q$. We therefore have
$\varepsilon =\varepsilon(s,q,\xi_1,h_1)$ with
\be
\delta\varepsilon =T\delta s+\mu \delta q-4\xi_2\delta \xi_1 -4h_2\delta h_1
\ee
and we can identify $\varepsilon$ as the energy density of the thermal system.

As a consistency check we can also calculate the energy by calculating the
holographic energy-momentum tensor. From \cite{bk} we have
\be
(8\pi G)T_{ij}=K_{ij}+\gamma_{ij}(-K+4+\tfrac{3}{2}\xi^2 +\tfrac{3}{2}h^2)
\ee
where $\gamma_{ij}$ is the spatial metric at a fixed radius $r$, $K_{ij}$ is the extrinsic curvature
tensor and we note that the $\xi$ and $h$ terms have arisen from the corresponding terms in $I_{ct}$
given in \reef{ctermact}.
We find that the $T_{ij}$ is diagonal with
\bea
T_t{}^t&=&\frac{1}{r^3}\frac{1}{2}(\varepsilon)\nn
T_x{}^x=T_y{}^y&=&-\frac{1}{r^3}\frac{1}{2}(\tfrac{1}{2}\varepsilon+{2}\xi_1 \xi_2+{2}h_1 h_2)
\eea
Using equation (45) of \cite{bk} to calculate the total energy  we obtain
\be
E=e^{\beta_a/2}\varepsilon vol_2
\ee
which agrees upon setting $\beta_a=0$ which we will do.
It is also worth noting that from the spatial part of the stress tensor
we deduce that the pressure is
$p=\tfrac{1}{2}\varepsilon + \tfrac{3}{2}\xi_1 \xi_2+\tfrac{3}{2}h_1 h_2$
and we thus see that the Smarr-type formula \reef{smarr} can be written
in the familiar form $\varepsilon+p=\mu q+ Ts$.

\section{Uncharged domain wall solutions: holographic RG flows}
Recall that the scalar potential for the truncated action \reef{lagred} with $\chi=\xi\in \bbR$ is given
by
\be
V = 24 \frac{-1 + h^2 + \xi^2}{(1-h^2)^{3/2}(1-\frac{3}{4}\xi^2)^2}
\ee
and has extrema for $(h,\xi)=(0,0), (0,\pm\sqrt{2/3}), (\pm 1/\sqrt{5},\pm 2/\sqrt{15})$.
These correspond to the skew-whiffed, Pope-Warner and Englert $AdS_4$ vacua, respectively, that were discussed in section 3.
As before, we will restrict our considerations to vacua with positive values of
the fields. Before discussing new solutions related to holographic superconductivity in subsequent sections,
in this section we pause to numerically construct interpolating uncharged domain-wall solutions,
with vanishing gauge field, $\phi=0$, which have the interpretation as ordinary holographic RG flows.

\begin{figure}[ht!]
\begin{center}
\includegraphics[width=0.45\textwidth]{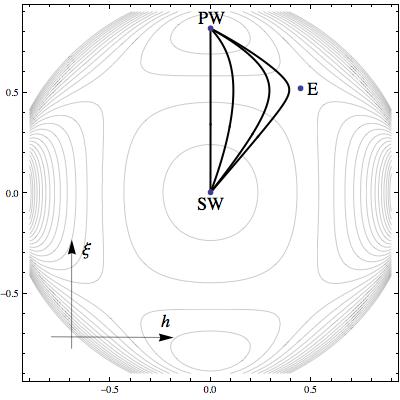}\caption{The plot shows the scalar potential of our model in the $(h,\xi)$ plane.
The extrema indicated by dots correspond to the skew-whiffed $AdS_4$ vacuum (SW), the Pope-Warner vacuum (PW)
and the Englert vacuum (E). The interpolating trajectories are domain wall solutions that describe holographic
flows interpolating between a deformed skew-whiffed vacuum in the UV and a Pope-Warner vacuum in the IR.\label{fig:scalarpot1}}
\end{center}\end{figure}

For these solutions,
as $r\to \infty$ we require the asymptotic expansion
given in \reef{aesw} (with $\mu=q=0$), corresponding to a perturbed skew-whiffed
vacuum in the UV.
For uncharged domain wall solutions that flow to the Pope-Warner vacuum in the IR, we demand that
as $r\to 0$ we have the expansion \reef{irexp} (with $\phi_{IR}=0$).
Note that in the UV $\xi,h$ are both dual to relevant operators and non-zero values of $\xi_1,h_1$
correspond to deforming the skew-whiffed CFT by the corresponding operators, ${\cal O}_h$ and ${\cal O}_\xi$, while
$\xi_2,h_2$ correspond to giving vevs for these operators in the deformed CFT.
By contrast, in the IR $\xi,h$ are both dual to irrelevant operators in the Pope-Warner
CFT.

Do such interpolating domain wall solutions exist? A simple counting
suggests the following picture. We have four fields, $g,\beta,\xi,h$, two of which
satisfy\footnote{Note that the domain walls we are interested in do not satisfy first order RG flow equations.
For example when $h=0$, after redefining $\xi=(2/{\sqrt 3})\tanh(s/2)$, the $D=4$ Lagrangian can be written
in the form
\be
16\pi G {\cal L}={\sqrt{-g}}[R-\tfrac{1}{2}(\nabla s)^2-V]
\ee
with the potential $V$ given in terms of a superpotential $W$ via $V=8(4(W')^2-3W^2)$ and $W=1/2(1+\cosh(s))$.
This is of the form considered in e.g. \cite{st} and we note that
while for the skew-whiffed vacuum $W'(0)=0$, by contrast
for the Pope-Warner vacuum $W'(1)\ne 0$.} first-order
equations and the rest second order equations (given in appendix A). Thus we must specify six constants to obtain
a unique solution. We can use the scaling symmetries of the theory given in
\reef{ss}, \reef{sv} to set $\beta_a=0$ as well as $\hfac\xi_1=1$. We therefore have seven
parameters, $\varepsilon, \xi_2, h_1, h_2, a_\beta, a_\xi, a_h$
left to specify and thus we expect to obtain a one-parameter family of solutions.
Let us take this parameter to be $h_1$. We will take $h_1\ge 0$ and solutions with
$h_1$ negative can be obtained using the $h\to -h$ symmetry of the equations of motion.

\begin{figure}[ht!]
\begin{center}
\includegraphics[width=0.45\textwidth]{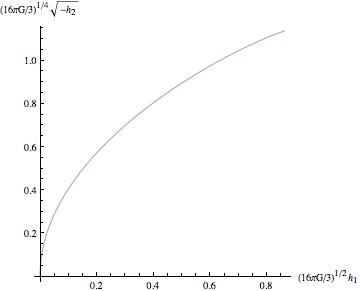}\quad \includegraphics[width=0.45\textwidth]{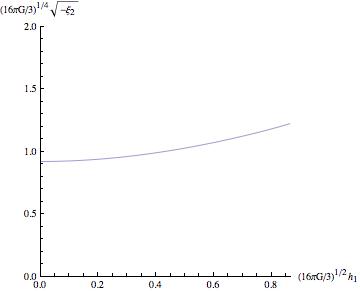}
\caption{Plot showing the behaviour of $h_2$ and $\xi_2$ as a function of $h_1$ in the
uncharged domain wall solutions.
\label{fig:neutralscan}}
\end{center}\end{figure}

Using a shooting technique (see Appendix \ref{app:num}), we have constructed this
one-parameter family of solutions numerically, finding solutions with $h_1$ in the range
$\hfac h_1\in [0,\sim .86)$.
Figure \ref{fig:scalarpot1} shows a contour plot of the scalar potential with
domain-wall trajectories superposed.
The solution with $h_1=0$ has $h=0$ identically.
As $h_1$ approaches the maximum value $\hfac h_1\sim 0.86$ the solution gets closer and closer to the unstable
Englert vacuum.
In Figure \ref{fig:neutralscan} we show the values of $h_2$ and
$\xi_2$, which are fixing the vevs $<{\cal O}_h>$ and $<{\cal O}_\xi>$, as a function of $h_1$.

Following similar considerations we have also constructed two further domain wall solutions, one that interpolates between the
skew-whiffed vacuum in the UV and the Englert vacuum in the IR and another between the
Englert vacuum in the UV and the Pope-Warner vacuum in the IR.
Since the Englert solution is unstable, the physical significance
of such solutions, if any, is not clear.

For the one parameter family of domain wall solutions flowing between
the skew-whiffed vacuum to the Pope Warner vacuum to describe sensible RG flows
we require that the corresponding $AdS_4$ solutions of $D=11$ supergravity are stable,
at least perturbatively. While perturbative stability has been demonstrated for the skew-whiffed solutions
it has not yet been shown for the Pope-Warner solutions.
Assuming that they are in fact stable, one might then be concerned that the instability of the
Englert vacuum implies a concomitant pathology of the RG flows, especially for values of $h_1$
near the maximum value $\hfac h_1\sim 0.86$ for which the domain wall solutions are getting close to the Englert solution.
We think that this is unlikely to be a problem. While the solutions do have a region that is approximated by
the Englert solution, the unstable mode of the Englert solution will not be localised in that region.
Furthermore, if there
was a critical value of $h_1$ for the solutions in which they become unstable, one would expect a marginal static
mode to appear which we are able to explicitly test for numerically, as described in detail in appendix \ref{app:num}, and do not find.

\section{Interpolating solutions with $T=0$, $\mu\ne 0$}
\label{intsol}
In this section we will study two classes of regular interpolating solutions with non-zero gauge field, $\phi\ne 0$,
that arise as the zero temperature limit of black hole solutions which will be constructed in section 9.
The first class is a one parameter family of charged domain walls that interpolate between deformed skew-whiffed $AdS_4$ in the UV and the
Pope-Warner $AdS_4$ solution in the IR. These solutions have scalar $\xi$ hair and, as we will show in section 9, are the zero temperature limit of superconducting black holes.
In particular, the $AdS_4$ region in the IR corresponds to an emergent $d=3$ conformal symmetry in the IR. For the special case when $h=0$ these solutions
were found in \cite{Gauntlett:2009dn}\cite{Gubser:2009gp}. The second class of solutions
is a one parameter family of charged solutions that interpolate between deformed skew-whiffed $AdS_4$ in the UV and $AdS_2\times\bbR^2$ in the IR. These
solutions have no scalar hair and, as we show later, will give the zero temperature limit of some of the normal phase black holes. For the special case when $h=0$ these
solutions are simply the zero temperature limit of the Reissner-Nordstr\"om black hole solution given in \reef{rnsol}. As we discuss in section 9, our numerical results indicate that the class of interpolating solutions with  $AdS_2$ factors in the IR are never thermodynamically favoured while those with $AdS_4$ factors are.


\subsection{Pope-Warner IR: zero temperature superconductors}

\begin{figure}[ht]
\begin{center}
\includegraphics[width=0.45\textwidth]{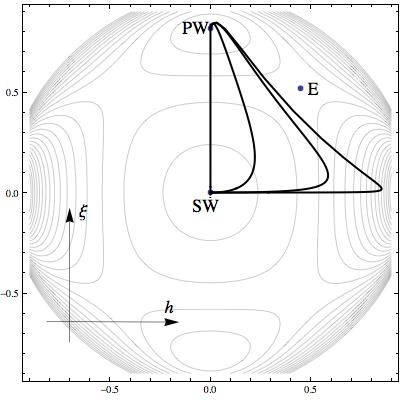}\caption{Plot in the $(h,\xi)$ plane
showing the interpolating trajectories
of the charged domain wall solutions interpolating between the skew-whiffed vacuum in the UV and the Pope-Warner
vacuum in the IR. Along the trajectories $\phi\neq0$.
\label{fig:scalarpot2}}
\end{center}\end{figure}

\begin{figure}[ht]
\begin{center}
\includegraphics[width=0.45\textwidth]{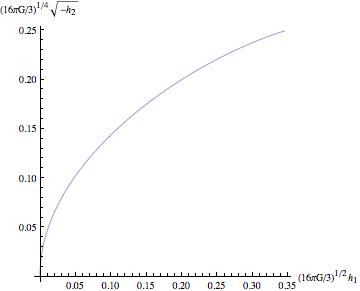}\\
 \includegraphics[width=0.45\textwidth]{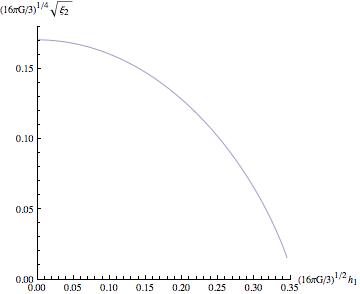}\quad \includegraphics[width=0.45\textwidth]{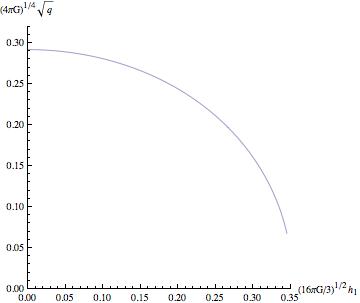}\caption{Plots showing $h_2$, $\xi_2$ and $q$ for the one-parameter family of charged domain wall solutions labelled by $h_1$ with $\mfac \mu=1$.    \label{fig:chargedscan}}
\end{center}\end{figure}

As $r\to \infty$
we again impose the asymptotic expansion given in \reef{aesw},
corresponding to a perturbed skew-whiffed vacuum in the UV. Similarly
as $r\to 0$ we impose the expansion \reef{irexp} in order that we approach the Pope-Warner
vacuum in the IR (observe that $\phi$ is dual to an irrelevant operator in the CFT dual
to the Pope-Warner vacuum). We now have five fields, $g,\beta,\xi,h,\phi$, two of which satisfy first-order
equations and the rest second order equations (given in appendix A). Thus we must specify eight constants to obtain
a unique solution. We next use the scaling symmetries of the theory given in
\reef{ss}, \reef{sv} to set $\mfac\mu=1$ as well as $\beta_a=0$. This leaves the ten parameters
$\varepsilon,q,\xi_1, \xi_2, h_1, h_2, a_\beta, a_\xi, a_h, a_\phi$ and so we expect a two
parameter family of solutions. We will fix one of these parameters by choosing to set $\xi_1=0$ (as we discuss further below).
This then leaves us with a single parameter which we choose to be $h_1$. Once again we take $h_1\ge 0$ and we can recover
negative $h_1$ by using the symmetry $h\to -h$.
Using a shooting techniques described in appendix C, we do indeed find a one parameter family of such charged domain wall solutions for
$h_1\le h_1^c$ with $\hfac h_1^c\sim 0.35$ (for $\mfac\mu=1$). In Figure
\ref{fig:scalarpot2} we have plotted the trajectories of the scalar fields and
Figure \ref{fig:chargedscan} displays the dependence of $h_2$, $\xi_2$ and $q$ on $h_1$. Notice that as $h\to h_1^c$, the charge carried by the black hole is going to zero since
$q\to 0$.

The special solution with $h_1=0$, which has $h=0$ identically, has been shown to
arise as the zero temperature limit of holographic superconducting black holes with
non-zero chemical potential in \cite{Gauntlett:2009dn}
We will see in the next section that all of the new charged domain walls arise in a similar way.
As in \cite{Gauntlett:2009dn} we have imposed $\xi_1=0$ because it corresponds to
allowing the operator ${\cal O}_\chi$, dual to $\xi$, to obtain a vev, determined by $\xi_2$, without being sourced
i.e. without adding the operator to the CFT dual to the skew-whiffed vacuum. Equivalently,
the abelian symmetry in the dual CFT is then broken spontaneously and not explicitly\footnote{Note that we expect entirely analogous results
if we set $\xi_2=0$ and $\xi_1\neq 0$.}.
In our new charged domain walls with $h\ne 0$ we necessarily have
$h_1\ne 0$, $h_2\ne 0$ corresponding to the dual $d=3$ CFT having been perturbed by the operator ${\cal O}_h$, dual to $h$, as well
as this operator acquiring a vev.
As we will discuss later the coupling of $h$ to $F\wedge F$ in \reef{lagred} implies
that when $h\ne0$ the dual theory breaks parity and time reversal invariance.

In the far IR, all of the new charged domain wall solutions approach the Pope-Warner CFT. Thus, in this limit,
all of the $h$-deformed skew-whiffed CFTs with $T=0$ and $\mu\ne 0$ are described by the same universal CFT.

\subsection{$AdS_2\times \mathbb{R}^2$ IR: zero temperature normal phase}

To construct the solutions interpolating between deformed skew-whiffed $AdS_4$ in the UV and $AdS_2\times \bbR^2$ in the IR, we proceed in a similar fashion.
However, since the scalar $\xi$ has no irrelevant behaviour about $AdS_2\times \mathbb{R}^2$ (see \reef{irexp2})
these solutions have $\xi=0$ identically (i.e. no scalar hair). The counting of parameters is exactly
the same as that for the Pope-Warner domain walls after excluding $\xi$, and so we again expect a one parameter family of solutions, which we label by $h_1\ge 0$.

\begin{figure}[th]
\begin{center}
 \includegraphics[width=0.45\textwidth]{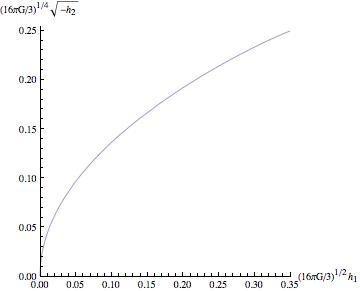}\quad \includegraphics[width=0.45\textwidth]{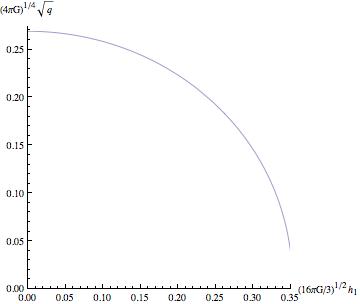}\caption{Plots showing $h_2$ and $q$ for the one-parameter family of charged domain wall solutions labelled by $h_1$ with $\mfac \mu =1$.
\label{fig:chargedscan2}}
\end{center}\end{figure}

We do indeed find a one parameter family of interpolating solutions which for  $h_1=0$ (in fact $h=0$ identically) includes the $T=0$ Reissner-Nordstr\"om solution \reef{rnsol}.
As we will see later all of these solutions arise as the zero temperature limit of charged black holes with $\xi=0$ corresponding to the normal
phase of the system. 
In Figure \ref{fig:chargedscan2} we have plotted the dependence of $h_2$ and $q$ on $h_1$. Note that as for the Pope-Warner charged domain walls,
as $h\to h_1^c$ it appears that the charge carried by the black hole is vanishing,
$q\to 0$.

Interestingly we find that these solutions exist for a very similar, if not identical, range of $h_1$ as for the Pope-Warner charged domain wall solutions of
the last section, i.e. $h_1\le h_1^c$ with $\hfac h_1^c\sim 0.35$ (for $\mfac\mu=1$). 
We will see later that the broken phase superconducting phase solutions are thermodynamically preferred when $h_1<h_1^c$. Thus, if the ranges are indeed identical (as we expect),  for $h_1<h_1^c$ the system at zero temperature 
is described by the iterpolating solutions with $AdS_4$ factors in the IR
and not those with $AdS_2$ factors. 
More detailed investigations near $h_1=h_1^c$ would certainly be worthwhile.

\section{Uncharged black hole solutions}

In this section we construct uncharged black hole solutions
that are asymptotic to the perturbed skew-whiffed $AdS_4$ solution.
These describe the skew-whiffed CFT deformed by the relevant operator ${\cal O}_h$ at finite temperature $T$ and $\mu=0$.
These solutions have unbroken gauge symmetry and we will see later how they interface with the superconducting solutions.

The asymptotic behaviour of the uncharged black holes is given in
\reef{aesw}
and the behaviour at the black hole horizon is given by \reef{exphor}.
The black hole solutions that we construct have $\xi=\phi=0$ identically,
but can have $h_1 \ne 0$, which corresponds to a deformation of the skew-whiffed CFT by
the operator dual to $h$, ${\cal O}_h$, as well as having $h_2\ne 0$ corresponding
to giving ${\cal O}_h$ a vev.
For solutions with $h_1=0$ (which have $h=0$ identically) we have
the usual neutral AdS-Schwarzschild black hole,
\be
g=4r^2-\frac{8 \pi G \varepsilon}{r}
\ee
As the temperature is taken to zero, one recovers the skew-whiffed $AdS_4$ vacuum in the usual manner.
For $h_1\ne0$ we have found new solutions for all temperatures.
To solve the differential equations we use the scaling symmetries \reef{ss}, \reef{sv}
to set $\beta_a=0$ and $\hfac h_1 = 1$. Since $\phi = \xi = 0$ we set $\mu =q=\xi_1 = \xi_2=0$.
This leaves us with five parameters $\varepsilon, h_2, r_+, \beta(r_+),  h(r_+)$ and since a solution to
the differential equations for $g,\beta,h$ is specified by four parameters we
expect a one parameter family of black hole solutions. We take this parameter to be the temperature of the black hole.

\begin{figure}[ht!]
\begin{center}
\includegraphics[width=0.45\textwidth]{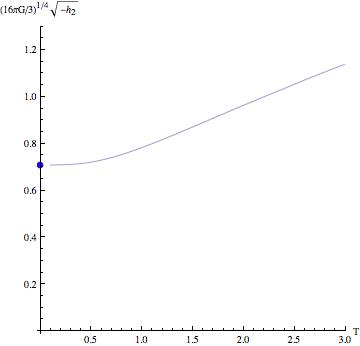}\qquad \includegraphics[width=0.45\textwidth]{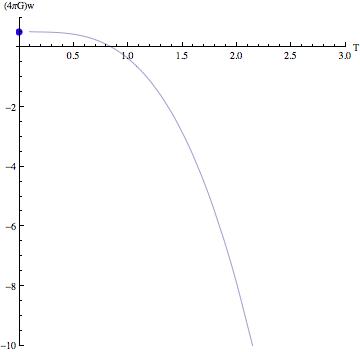}\\
\includegraphics[width=0.45\textwidth]{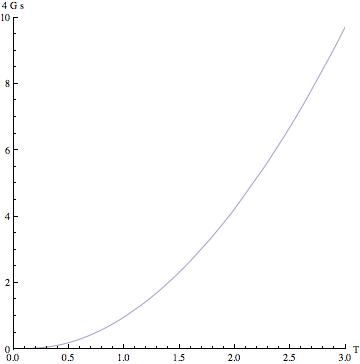}
\caption{Plot showing the dependence of $h_2$, the thermodynamic potential $w$ and
the entropy density $s$ 
against temperature for the uncharged black hole solutions with $\hfac h_1 = 1$.
The dots are explained in the text.
\label{unchargedh2}}
\end{center}\end{figure}

\begin{figure}[ht!]
\begin{center}
\includegraphics[width=0.4\textwidth]{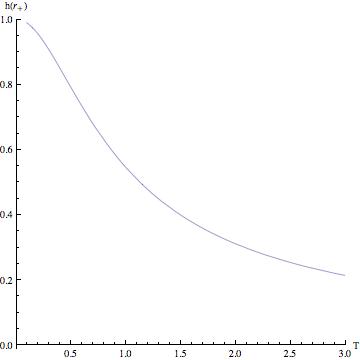}\quad \includegraphics[width=0.4\textwidth]{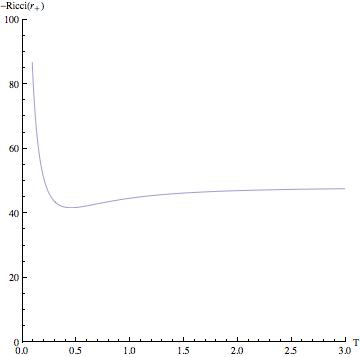}\\
 \caption{Plot showing the value of the $h$ and the Ricci scalar at the horizon against temperature
 for the uncharged black hole solutions with $\hfac h_1 = 1$.
\label{unchargedRicci}}
\end{center}\end{figure}

In Figure \ref{unchargedh2} we have plotted the dependence of $h_2$, the thermodynamic
potential $w=-\varepsilon/2-2h_1h_2$ (since $\xi_1=0$) and also
$r_+^2=4Gs$, where $s$ is the entropy density, against
temperature. Observe that as the temperature goes to zero the entropy goes to zero.
In Figure \ref{unchargedRicci} we have plotted the value of $h$ at the horizon, $h(r_+)$ and
we see that as the temperature goes to zero it is approaching the singular value of $1$. We
have also plotted the Ricci scalar at the horizon as a function of temperature
in Figure \ref{unchargedRicci} which confirms that,
unlike the solutions with $h_1 = 0$, as the temperature is decreased to zero, the
solutions become singular.

We have verified that as the temperature goes to zero, the solution appears to approach Poincar\'e invariant
behaviour with $ge^{-\beta}=r^2$. First observe that it is consistent with the equations of motion to set
$ge^{-\beta}=r^2$ when $\phi=\xi=0$ and that the equations then boil down to solving a second order ODE for $h$. 
We find that as $T\to 0$ our solutions approach the behaviour $1-h\sim r^{4/3}$ and hence $g\sim r^{4/3}$ near the singularity at $r=0$. 
Note that this behaviour implies that the distance to the singularity from any fiducical point in the spacetime is finite.
In fact we have found a full analytic solution of the second order ODE for $h$ given by
\bea
h & = & \frac{\sqrt{3} \sqrt{\frac{6 \sqrt{3} \left(2 r^8+4 r^4+1\right)}{\sqrt{Z}}-Z+9}-\sqrt{3} \sqrt{Z}+3}{6 \left(r^4+1\right)} \nonumber \\
Z & = & \frac{2 \; 6^{2/3} \left(r^4+1\right) r^{8/3}}{  \left(  \sqrt{48 r^4+81}-9  \right)^{1/3}  }
-{6}^{1/3} \left(r^4+1\right) \left( \sqrt{48 r^4+81}-9 \right)^{1/3} r^{4/3}+3 
\eea
which appears to describe the $T\to 0$ solution. As $r\to 0$ it has the behaviour
$h = 1 - \frac{1}{2^{1/3}} r^{4/3}=\dots$, while as $r\to \infty$,
$h = 1 / r - 1 /(2 r^2 ) + ...$, i.e. $\hfac h_1 = 1,  \hfac h_2 = - 1/2$ and 
$g = 4 r^2 + 3  - 4/r + ...$ i.e. $4\pi G\varepsilon  
=4\pi G w=1/2$. We have indicated this behaviour on Figure \ref{unchargedh2} with dots for comparison.

\section{Charged black hole solutions and superconductivity}

In this section we construct charged black hole solutions both without and with charged scalar hair that describe
unbroken and superconducting phases of holographic superconductors, respectively.
We also connect the zero temperature limit of these black holes to the interpolating solutions discussed in section \ref{intsol}.
Finally, we describe our calculations on the electrical conductivity of the black holes.

The asymptotic behaviour of the charged black holes is given in
\reef{aesw}, corresponding to a perturbed skew-whiffed vacuum, and
the behaviour at the horizon is given by \reef{exphor}.
The high temperature unbroken phase black hole solutions that we construct have $\xi=0$ identically
but can have $h_1\ne 0$, which corresponds to a deformation of the skew-whiffed CFT by
the operator dual to $h$, ${\cal O}_h$, as well as having $h_2\ne 0$ corresponding
to giving ${\cal O}_h$ a vev. On the other hand
the charged black hole solutions with charged scalar hair, corresponding to the low
temperature superconducting phase,
will have $\xi_1=0$ and $\xi_2\ne 0$, corresponding to allowing the
operator dual to $\xi$ in the skew-whiffed CFT
to acquire a vev without being sourced\footnote{Note that as for the charged domain walls discussed in section 7.1, we again expect analogous results
if we set $\xi_2=0$ and $\xi_1\neq 0$.}.
These black hole solutions will also have, generically,
$h_1,h_2 \ne0$.

To solve the differential equations we first use the scaling symmetries \reef{ss}, \reef{sv}
to set $\beta_a=0$ and $\mfac \mu =1$. We also impose $\xi_1=0$. This leaves us with ten parameters $\varepsilon,q,\xi_2, h_1, h_2$, $\beta(r_+), \xi(r_+), h(r_+), \phi'({r_+})$ and $r_+$, and since a solution to
the differential equations for $g,\beta,\xi,h,\phi$ is specified by eight parameters we
expect a two-parameter family of black hole solutions. We can choose these parameters to be the temperature $T$ and $h_1\ge 0$ (with negative $h_1$ obtained via $h\to -h$).
For further discussion see appendix C.
It is worth noting that when $T\to 0$ we have the same parameter counting that arose in the charged domain-wall
solutions discussed in section \ref{intsol}.1. This is because the operators dual to
$\xi,h$ and $\phi$ are all irrelevant in the Pope-Warner CFT.
When $h_1=0$ the solutions have $h=0$ identically and were all constructed in \cite{Gauntlett:2009dn}.

\subsection{Black hole solutions with $\mu\ne 0$ and no scalar hair: the unbroken phase}

The unbroken phase black hole solutions have $\xi=0$ identically.
When $h_1=0$ (in fact $h=0$ identically) we have
the usual Reissner-Nordstr\"om kind of electrically charged black hole solutions given in \reef{rnsol}.
Recall that at zero temperature the Reissner-Nordstr\"om solutions have finite entropy and
near the event horizon approach $AdS_2\times \bbR^2$ with the $AdS_2$ having radius squared $1/{24}$
and hence the Ricci scalar is -48.
For non-zero values of $h_1$ with $h_1<h_1^c$, with $\hfac h_1^c\sim 0.35$ (for $\mfac\mu=1$), we
have found solutions with very similar properties. In particular, at zero temperature they
approach the interpolating solutions presented in section 7.2 that approach $AdS_2\times \bbR^2$ in the IR.
As we will see below, this is the same range of $h_1$ for which superconducting black hole solutions also exist.
We have also constructed charged unbroken phase solutions for $h_1>h_1^c$ and for this range the solutions become singular
in the zero temperature limit.

\begin{figure}[t!]
\begin{center}
\includegraphics[width=0.45\textwidth]{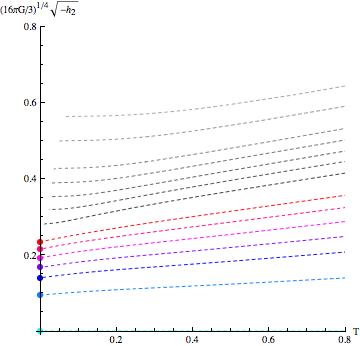}\hskip2em\includegraphics[width=0.45\textwidth]{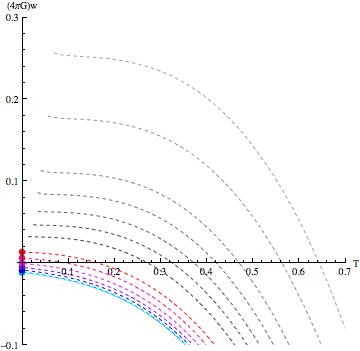}
\end{center}

\begin{center}
\includegraphics[width=0.45\textwidth]{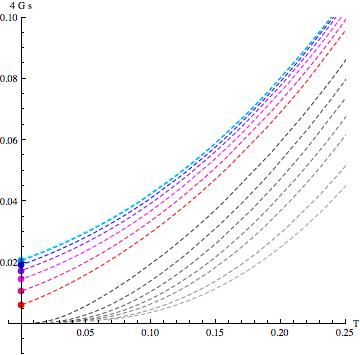}
\caption{Plot showing the dependence of $h_2$,
the thermodynamic potential $w$ and entropy density $s$
against temperature for the unbroken phase black hole solutions
for various values
of $h_1$ ranging from $\hfac h_1=0$ (light blue) to $\hfac h_1=0.3$ (red).
Increasing values of $\hfac h_1=0.4,\dots, 0.8$ are shown in decreasing shades of
grey. The zero charged domain wall solutions are added at $T=0$ as
coloured dots. All plots have $\mfac\mu=1$. \label{h2}}
\end{center}\end{figure}

\begin{figure}[th]
\begin{center}
\includegraphics[width=0.45\textwidth]{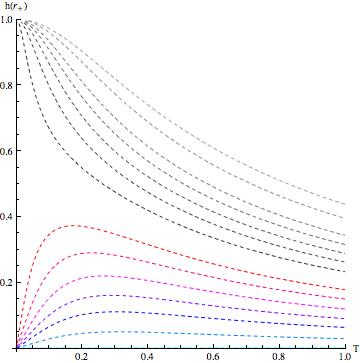}\quad \includegraphics[width=0.45\textwidth]{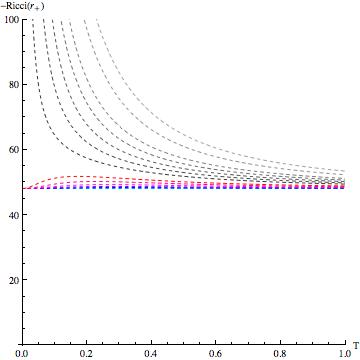}\\
 \caption{Plots showing the value of $h$ and of the Ricci scalar at the horizon against temperature
 for the unbroken black hole solutions for various values
for various values
of $h_1$ ranging from $\hfac h_1=0$ (light blue) to $\hfac h_1=0.3$ (red).
Increasing values of $\hfac h_1=0.4,\dots, 0.8$ are shown in decreasing shades of
grey. The zero charged domain wall solutions are added at $T=0$ as
coloured dots. Both plots have $\mfac\mu=1$.
\label{upbh}}
\end{center}\end{figure}

In Figure \ref{h2} we have plotted the dependence of $h_2$, $r_+^2=4Gs$, where $s$ is the entropy density,
and also the thermodynamic potential $w=-\varepsilon/2-2h_1h_2$ (since $\xi_1=0$) against
temperature for various values of $h_1$. In Figure \ref{upbh} we have plotted the value of $h$ and the value of the Ricci scalar
at the horizon $r=r_+$ against temperature, which clearly demonstrates the change from non-singular to singular behaviour
at zero temperature as $h_1$ becomes bigger than $h_1^c$.

\subsection{Black hole solutions with $\mu\ne 0$ and scalar hair: the superconducting phase}

\begin{figure}[t!]
\begin{center}
\includegraphics[width=0.45\textwidth]{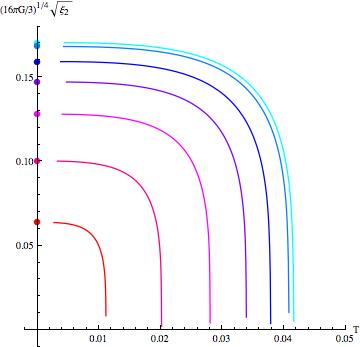}\qquad
\includegraphics[width=0.45\textwidth]{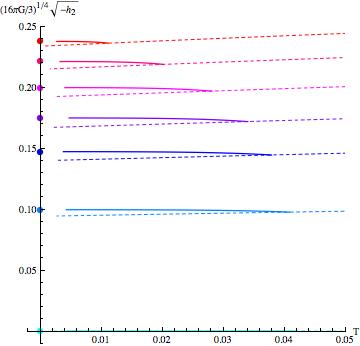}\\
\includegraphics[width=0.45\textwidth]{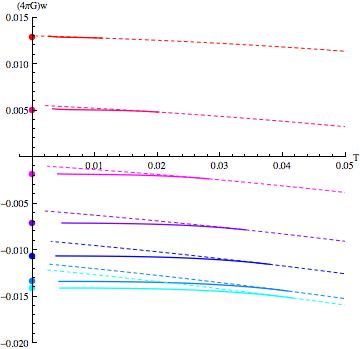}
\caption{Plot showing dependence of $\xi_2$, $h_2$ and the thermodynamic potential $w$ with
temperature for the broken phase black hole solutions for various values
of $h_1$ ranging from $\hfac h_1=0$ (light blue) to $\hfac h_1=0.3$ (red)
with $\mfac \mu=1$.
These broken phase solutions are plotted as solid lines and
the zero temperature charged domain wall solutions are added at $T=0$ as coloured dots.
The unbroken phase black solutions for the same values of $h_1$ are plotted as dashed lines for comparison.
\label{sig}}
\end{center}\end{figure}

We have constructed charged black holes with charged scalar hair, for various values of $h_1$, with $h_1<h_1^c$ where $\hfac h_1^c\sim 0.35$
(for $\mfac\mu =1$). As we have already noted this is the same range of $h_1$ that is required for the existence of
a zero temperature charged domain wall solution that we constructed in section 7.1.
In Figure \ref{sig}
we have plotted $h_2$, $\xi_2$ (recall $\xi_1=0$ for these solutions) and $w$ against temperature for various values
of $h_1$ for these new solutions\footnote{To compare with Figures 1 and 2 in \cite{Gauntlett:2009dn} which plots the solutions for $h=0$ one should note that
the normalisation in \cite{Gauntlett:2009dn} was $\mfac\hat\mu=1$, as we are using here. In addition the vertical axis in Figure 1
of \cite{Gauntlett:2009dn} was missing a factor of $4\pi$.}.
Importantly Figure \ref{sig} shows that the superconducting phase solutions, with charged scalar hair,
are thermodynamically favoured over the unbroken phase solutions.
Our numerical results also clearly indicate that as the temperature goes to zero, the broken phase black hole solutions do indeed
approach the charged domain wall solutions that we constructed in section 7.1. We have made various checks of this including
comparing scalar curvature invariants as was done for $h=0$ in \cite{Gauntlett:2009dn}.
In particular, we find that as the temperature goes to zero the entropy of the superconductors goes to zero and that the solutions
have an emergent conformally invariant behaviour in the far IR that is
captured by the Pope-Warner $AdS_4$ vacuum for all values of $h<h_1^c$.

These results are consistent with the simple picture that the CFT deformed by the operator
${\cal O}_h$ and held at finite chemical potential $\mu\ne 0$ has a phase diagram as summarised in Figure 1.
For $h_1<h_1^c$ there is high temperature phase in which the abelian symmetry is unbroken and then below a critical temperature there is a superconducting phase in which it is broken.
At low temperatures, and in the far IR, the superconductors are dual to the Pope-Warner $AdS_4$ solution.
For $h_1>h_1^c$ there is only an unbroken phase and at low temperatures and in the far IR the system is dual to a singular solution.

To demonstrate that this is indeed the correct picture would require showing that there are no other
black hole solutions which branch off at temperatures higher than the critical temperatures that we have found and that also dominate the free energy.
Within our radial truncation, as we discuss in appendix C, we find no evidence for any such additional solutions.
It is possible that there are such  black hole solutions of $D=11$ supergravity outside our consistent KK truncation \reef{lagred};
perhaps it can be shown, at least for a class of $SE_7$ spaces, that additional black hole solutions do not exist, but we leave that for future work.

\subsection{Conductivity}
Linear response theory can be used to determine the various
conductivities of the dual field theories for both the
unbroken and broken phase black hole solutions.
We will consider small perturbations
of the form
\bea
g_{tx} &=& e^{-i\omega t} g_{tx}(r)\,,\qquad
g_{ty} = e^{-i\omega t} g_{ty}(r)\,,\nn
\delta A&=&e^{-i\omega t} [a_x(r)dx + a_y(r) dy]\nn
&=& e^{-i\omega t} [a_-(r)(dx+idy) + a_+(r)(dx-idy)]
\eea
where we have introduced the circularly polarised perturbations
\be
a_\pm(r)\equiv\frac{1}{2}[a_x(r)\pm i a_y(r)]
\ee
All other fields take their background values.

The perturbations behave in the UV as
\bea\label{eq:vectorUV}
a_i(r) &\sim& \sqrt{4\pi G}\left[a^{(0)}_i + \frac{a^{(1)}_i}{r}\right]\,,\qquad g_{ti}\sim g_{ti}^{(0)}r^2 + \frac{g_{ti}^{(1)}}{r}
\eea
where we have used the notation $(x,y)\equiv (x^1,x^2)$.
At finite temperature, where there is a black-hole horizon, we impose purely ingoing boundary conditions, with the perturbations behaving
as $(r-r_+)^{-i\omega/4\pi T}$. For the superconducting black holes in the zero temperature limit,
we impose similar ingoing boundary conditions at the IR end of the domain-wall geometry.
As explained in \cite{Hartnoll:2009sz} the above perturbation produces an electric field given by $E_i=i\omega (a^{(0)}_i+\mu g^{(0)}_{ti})$
while the induced current $J_i$ is obtained by calculating the on-shell variation of the action and we find $\delta S/\delta a^{(0)}_i=4a^{(1)}_i$.
Thus the electric conductivity matrix is given by
\be
\sigma_{ij} = \frac{-i 4a_i^{(1)}}{\omega a_j^{(0)}}
\ee
The rotational invariance of our setup implies that
$\sigma_{ij}=s_1\delta_{ij}+s_2\epsilon_{ij}$ and that $s_2$ is non-zero
only if parity and time reversal invariance is violated. As we will see when $h\ne 0$ the coupling to $F\wedge F$
in \reef{lagred} gives rise to such parity violation.
Note that the thermal $\bar\kappa$ and thermoelectric $\alpha$ conductivities can be easily determined once $\sigma$
is known \cite{Hartnoll:2007ip}\cite{Hartnoll:2009sz} .

The $xr$ and $yr$ Einstein equations imply the first order equations
\bea\label{eq:firstorderpert}
r^2 \left( \frac{g_{ti}}{r^2} \right)' =-\frac{4 (1-h^2)^{3/2}}{1 + 3 h^2} \phi' a_i\,\qquad\, i = x,y
\eea
We also find that the Maxwell equations for $a_\pm$ decouple. After using
\reef{eq:firstorderpert} and also \reef{phieqe}, we deduce that
\be\label{plusminus}
a_\pm'' + \left[ \frac{g'}{g} - \frac{\beta'}{2} + \frac{L_1'}{L_1} \right]a_\pm' +
\left[ \left(\frac{\omega^2}{g^2} - \frac{4\phi'^2L_1}{g}\right)e^\beta - \frac{12 e^{-12U}\xi^2}{L_1 g}
\pm \frac{3L_3\omega h' e^{\beta/2}}{L_1 g}\right]a_\pm=0
\ee
where
\be
L_1=\frac{(1-h^2)^{3/2}}{1+3h^2},\qquad L_3=\frac{(1-h^2)^2}{(1+3h^2)^2}
\ee
The only other non-zero components of the Einstein equations are in
the $xt$ and $yt$ components and they lead to second order equations which are implied by the
above.

Thus the problem boils down to solving the linear equations \reef{plusminus}
subject to the boundary conditions mentioned above.
The parity operation in the three-dimensional boundary field theory can be defined as the
reversal of one of the spatial directions, $x$ say, and this transforms
$a_\pm\to -a_\pm$. On the other hand time reversal changes the sign of $\omega$.
Thus, the last term in \reef{plusminus} implies that
configurations with non-constant $h$ break parity and time-reversal invariance
of the boundary field theory.

Since the equations for $a^\pm$ decouple, the conductivity matrix is diagonal in
this basis. We first calculate
\be
\sigma^{\pm} \equiv  \frac{-i 4a_{\pm}^{(1)}}{\omega a_\pm^{(0)}}
\ee
which can be used to obtain
\bea
\sigma_{xx}=\sigma_{yy}=\tfrac{1}{2}(\sigma^++\sigma^-)\nn
\sigma_{xy}=-\sigma_{yx}=\tfrac{i}{2}(\sigma^+-\sigma^-)
\eea
In particular, the special solutions with $h=0$ found in \cite{Gauntlett:2009dn}
have vanishing Hall conductivity $\sigma_{xy}=0$ but the new solutions found here with
$h\ne 0$ have $\sigma_{xy} \ne 0$.

\begin{figure}[th]
\begin{center}
\includegraphics[width=0.45\textwidth]{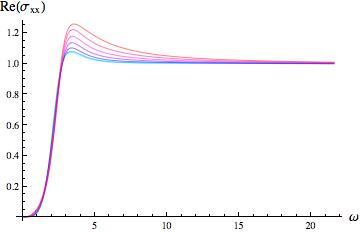}\qquad\includegraphics[width=0.45\textwidth]{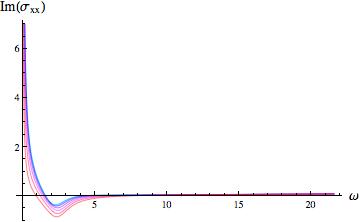}
\caption{The left panel shows the real part of the direct conductivity of the superconducting solutions at zero temperature for $\hfac h_1=0, 0..05,0.1,0.15,0.2,0.25,0.3$ (from bottom to top)
and $\mfac \mu=1$.
The right panel shows the corresponding imaginary part (now from top to bottom). We see that there is a pole in the imaginary part, at $\omega=0$, from which we deduce by analyticity
that the real part has a delta function there, which the numerics don't show.\label{fig:directcond}}
\end{center}\end{figure}

\begin{figure}[th]
\begin{center}
\includegraphics[width=0.45\textwidth]{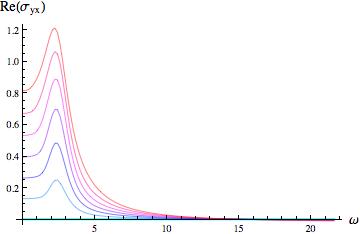}\qquad\includegraphics[width=0.45\textwidth]{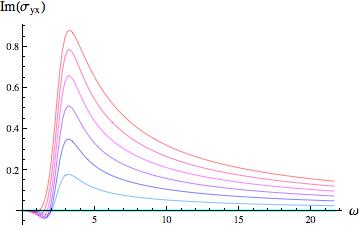}
\caption{The left panel shows the real part of the Hall conductivity of the superconductors at zero temperature for $\sqrt{16 \pi G/3}h_1=0, 0.05,0.1,0.15,0.2,0.25,0.3$ (from bottom to top)
and $\mfac \mu=1$.
The right panel shows the corresponding imaginary part (again from bottom to top). Note that the undeformed ($h_1=0$) Hall conductivity, shown in cyan, vanishes identically
because P and T remain unbroken at $h_1=0$.
\label{fig:Hallcond}}
\end{center}\end{figure}

We have calculated the electrical conductivity for various values of $h_1$ and for various temperatures for
both broken and unbroken phase black holes. Here we will just plot
the real and imaginary parts of $\sigma_{xx}$ and $\sigma_{xy}$ for the zero temperature superconducting charged domain wall solutions.
The longitudinal conductivity
is shown in Figure \ref{fig:directcond} and the Hall conductivity in Figure \ref{fig:Hallcond}.
At finite temperature we find similar results.

\section{Discussion}

Building on the results of \cite{Gauntlett:2009zw}, we have presented a simple consistent KK truncation
of $D=11$ supergravity on an arbitrary seven-dimensional Sasaki-Einstein space, $SE_7$,
to a $D=4$ theory that contains a metric, a gauge-field, a charged scalar field $\chi$ and a neutral scalar field $h$.
Given any $SE_7$, each solution of the $D=4$ theory gives rise to an exact
solution of $D=11$ supergravity.

\begin{figure}[ht!]
\begin{center}
\includegraphics[width=0.6\textwidth]{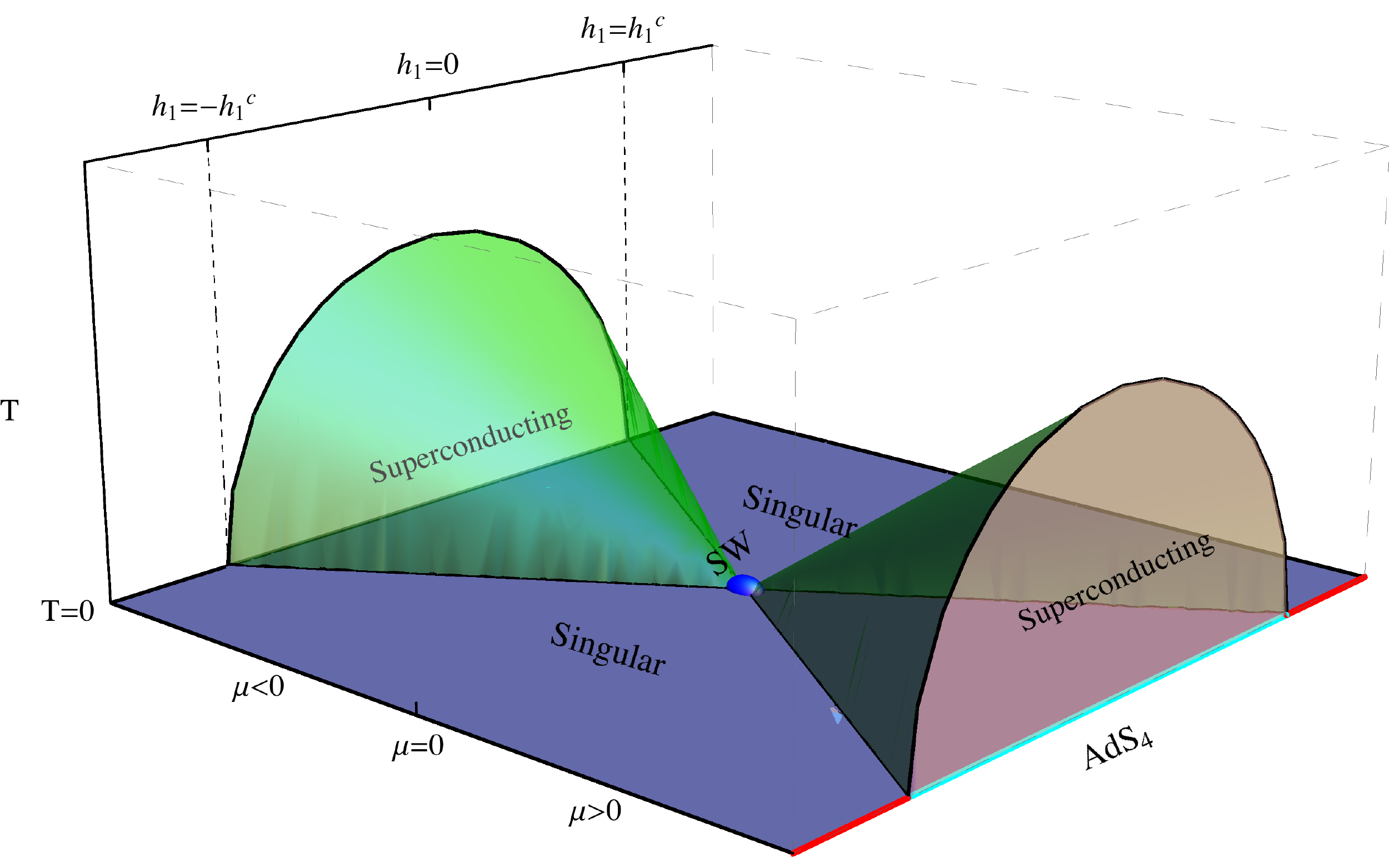}
\end{center}
\caption{The phase diagram of the holographic superconductors in the $\left( T,\mu,h_1 \right)$ space.  Note any $\mu\ne 0$ slice gives the $2d$ projection of Figure 1. At the origin we have the skew-whiffed vacuum. The $T=0$ plane is null singular except under the two half cones. \label{fig:3dphase}}
\end{figure}

We have used this $D=4$ theory to construct a range of solutions that describe the thermodynamics of
the CFT dual to the skew-whiffed $AdS_4$ vacuum, allowing for finite temperature, $T$,
finite chemical potential $\mu$ with respect to the current dual to the gauge-field, and also deformations
by the relevant operator ${\cal O}_h$ dual to the field $h$. This latter deformation is encoded in the asymptotic behaviour of the $h$-field: $h=\hfac(h_1/r+\dots)$.
The phase diagram is summarised in Figure \ref{fig:3dphase} and extends the results of \cite{Gauntlett:2009dn} which constructed the solutions with $h=0$.

When $\mu=0$, we constructed new uncharged black hole solutions in section 8 for $T\ne 0$ and $h_1\ne 0$. As $T\to 0$ the entropy of the solutions
goes to zero, in accordance with the third law, and they become nakedly singular. Interestingly, in this limit they
develop a Poincar\'e symmetry and furthermore the singularity is at finite proper distance from any fiducial point in the bulk of
the spacetime; it would be interesting to see if this leads to a gapped spectrum.  For the solutions with $h_1=0$ (which have $h=0$), the black hole solutions are the usual Schwarzschild $AdS$ solutions and smoothly map onto the skew-whiffed $AdS_4$ vacuum at $T=0$ as indicated in  Figure \ref{fig:3dphase}.

Richer structures appear when $\mu\ne 0$. We first point out that because of the underlying conformal symmetry the physics is captured by any fixed slice
$\\mu\ne 0$  as presented in Figure 1. In section 9 we constructed electrically charged black holes with vanishing scalar hair $\chi=0$ which describe
the unbroken phase of the system. For $|h_1|<h_1^c$ we showed that below a critical temperature there is a transition to a superconducting phase described
by electrically charged black holes with scalar hair $\chi\ne 0$. The region of superconductivity corresponds to the region below the two green half cones in Figure \ref{fig:3dphase}.
At zero temperature the superconducting solutions are the non-singular charged domain wall
solutions interpolating between the skew-whiffed $AdS_4$ vacuum and the Pope-Warner $AdS_4$ vacuum constructed in section 7.
It is interesting that in the far IR these solutions have emergent conformal symmetry in three spacetime dimensions, described by the same CFT.

At $|h_1|=h_1^c$ the critical temperature goes to zero, and for $|h_1|>h_1^c$
the unbroken phase black hole solutions extend all the way down to zero temperature.
In the zero temperature limit these solutions again have vanishing entropy and become nakedly singular;
it would be interesting to explore them in more detail.
Interestingly, in all of these solutions with $h\ne 0$, parity and time reversal
invariance are broken in the boundary theory.

A striking aspect of our numerical investigations is that $|h_1|=h_1^c$ also seems to precisely correspond to
a change of behaviour of the zero temperature limit of the unbroken phase black holes. 
In particular, for $|h_1|<h_1^c$ at zero temperature these solutions become smooth interpolating solutions
that approach $AdS_2\times \bbR^2$ in the IR and are thermodynamically disfavoured.
It would be worthwhile further exploring both the zero temperature superconducting black holes and the unbroken phase black holes near
their critical values of $h_1$ to confirm that $h_1^c$ is indeed governing both behaviours and also to elucidate what
is physically happening precisely at $|h_1|=h_1^c$.

For simplicity, in all of our solutions that carry scalar charge we only considered asymptotic behaviours
of $\hfac \xi=\xi_1/r+\xi_2/r^2+\dots$ with $\xi_1=0$ and $\xi_2\ne 0$ corresponding to 
a spontaneous breaking of the abelian gauge symmetry by the operator dual to $\xi$ acquiring a vev. All of our solutions should have analogues
with $\xi_1\ne 0$ and $\xi_2= 0$, describing the same phenomenon in the CFT dual to skew-whiffed $AdS_4$ but
with different boundary conditions.
We do not anticipate any significant qualitative differences in behaviour but we think it would be worthwhile to investigate this in detail.

Given the rich phase structure that we have uncovered in this work, one may ask how this 
might map onto the phase diagram of a real material exhibiting
quantum critical behaviour. One may define a quantum critical point (QCP) to be a zero-temperature 
second order phase transition with emergent scaling symmetry, and we will
restrict our discussion here to cases where this is enlarged to a conformal symmetry. In the vicinity of such a QCP a condensed-matter system,
which typically has a discrete underlying lattice structure, is well-described by a relativistic continuum CFT. The coupling that one has to dial in order to
reach the quantum critical point shows up as a relevant deformation in the CFT.
Real-word examples include the applied pressure in heavy-fermion systems and also, conjecturally,
the hole-doping ratio in the cuprate superconductors.

The simplest possibility is that 
the skew-whiffed CFT, dual to the skew-whiffed $AdS_4$ vacuum (for a given choice of $SE_7$ space),
provides an AdS/CFT realization of a material with such a QCP.
Specifically, when $\mu=0$ we observe a phase transition at zero temperature when $h_1=0$.
In particular, tuning $h_1$ corresponds to tuning the operator ${\cal O}_{h}$, 
which is a relevant operator in the skew-whiffed CFT, and so one might try and identify ${\cal O}_{h}$
with some macroscopic quantity in the material.
When we switch on $\mu\ne 0$, at zero temperature the phase transition opens up into the two 
phase transitions at either end of the superconducting dome (see Figures 1 and 14) at $h_1 = \pm h_{1,c}$.
Interestingly underneath the superconducting dome at zero temperature there is
an emergent conformal scaling behaviour in the far IR (with the scale set by $\mu$), 
which is a universal strongly coupled CFT dual to the Pope-Warner $AdS_4$ vacuum. It would be very interesting if
there are real world materials with QCP's with such a phase structure.

It might be too optimistic to be able to find a material with a QCP that is 
exactly described by the skew-whiffed CFT for some choice of $SE_7$. One might then look for some approximate features.
Alternatively, one might also wonder if it is instead possible to interpret the CFT dual 
to the Pope-Warner $AdS_4$ vacuum as {\it the} QCP.
In this point of view, the IR description of the material near its QCP would coincide with the 
Pope-Warner CFT that the skew-whiffed CFT flows to in the far IR.
Thus, in this interpretation the skew-whiffed CFT would be a fictitious UV theory, 
whose usefulness for the real material lies solely in
the fact that at low energies it flows to an IR fixed point that describes the quantum critical region of that material.
One problem with this point of view is that, within our truncation, the PW fixed point has no relevant operators
and so one cannot deform away from it, as one could with the skew-whiffed CFT. 
This could potentially be solved by going outside the truncation used in this paper,
where one might find the required relevant operator.
We feel this warrants further investigation.

We conclude by briefly mentioning some additional avenues for further exploration. 
The first issue concerns the stability of the skew-whiffed $AdS_4$ solutions. 
For the special case that $SE_7=S^7$ the stability is guaranteed by supersymmetry. For the generic class, it is known
that they are perturbatively stable \cite{Duff:1984sv} 
and hence that they define CFTs in the strict $N\to \infty$ limit. It would be 
interesting to further study whether or not they are destabilised by $1/N$ effects of the type discussed in
\cite{Berkooz:1998qp}\cite{am}. It would also be interesting to have an explicit construction of the CFTs
dual to the skew-whiffed $AdS_4\times SE_7$ solutions.
Since the Pope-Warner solutions play an important role in our solutions,
it would also be worthwhile analysing their stability, starting with a 
perturbative analysis. 

Another outstanding issue is to determine whether or not there are additional unstable modes in the 
skew-whiffed $AdS_4\times SE_7$
background which condense at higher temperatures.
If they do exist, and dominate the free energy then the corresponding black hole solutions
would be the relevant ones for describing the thermodynamics. It would 
be nice to show, perhaps for a subclass of Sasaki-Einstein manifolds, that this does not occur.
In addtion to further explorations of superconductivity we envisage 
that our $D=4$ action will also have other applications in studying condensed matter systems in the context of M-theory.

\subsection*{Acknowledgements}
\noindent
We would like to thank Nigel Cooper, Sophia Domokos, Sean Hartnoll, Matt Headrick, Chris Herzog, Steve Gubser, Thomas Hertog, Igor Klebanov, Finn Larsen, Arttu Rajantie,
Ashoke Sen, Cobi Sonnenschein and Daniel Waldram for useful discussions.
J.P.G. is supported by an EPSRC Senior Fellowship and a Royal Society Wolfson Award,
J.S. by an EPSRC Postdoctoral Fellowship and
T.W. by an STFC Advanced Fellowship and a Halliday Award.
J.S. would like to thank the EFI, University of Chicago and the MCTP, University of Michigan for hospitality.

\appendix

\section{The differential equations}
Starting with the equations of motion arising from \reef{lagred} and substituting
in the ansatz \reef{mans}, \reef{gans}, \reef{sans}
we obtain differential equations for $g,\beta,\phi,\xi,h$.
The following differential equations arise from the matter fields:
\bea\label{phieqe}
\left[e^{\B/2}\R^2\frac{(1-h^2)^{3/2}}{1+3h^2}\phi'\right]'
-\frac{12\G^{-1}e^{\B/2}\R^2\phi\xi^2}{(1-\tfrac{3}{4}\xi^2)^2 }=0
\eea
\bea\label{rhoeqe}
&&r^{-2}e^{\beta/2}(1-\tfrac{3}{4}\xi^2) 
\left[\frac{\G e^{-\B/2}\R^2\xi'}{(1-\tfrac{3}{4}\xi^2)^2 }\right]'\nn
&&+\left[
\frac{9}{2}g(\xi')^2-\frac{12}{(1-h^2)^{3/2}}(-2+6h^2+3\xi^2)
+12g^{-1}e^{\beta}\phi^2(4+3\xi^2)
\right]\xi
=0
\eea
\bea\label{heqe}
r^{-2}e^{\beta/2}\left[\frac{r^2e^{-\beta/2}g}{(1-h^2)}h'\right]'
-\frac{2h(1-h^2)^{3/2}(3+h^2)}{(1+3h^2)^2}e^{\beta}(\phi')^2
-\frac{8h(-1+h^2+3\xi^2)}{(1-\tfrac{3}{4}\xi^2)^2 (1-h^2)^{3/2}}=0\nn
\eea

For the components of the Einstein tensor we have
\bea
E_{tt}&=&-\G^2e^{-\B}(\frac{\G'}{\G\R}+\frac{1}{\R^2})\nn
E_{\R\R}&=&\frac{\G'}{\G\R}-\frac{\B'}{\R}+\frac{1}{\R^2}\nn
E_{xx}&=&\frac{\R}{2}\left[2\G'-\G\B'\right]+
\frac{\R^2}{2}\left[\G''-\tfrac{3}{2}\G'\B'-\G\B''+\tfrac{1}{2}\G(\B')^2\right]
\eea
The $tt$ and $\R\R$ components of the Einstein equations are equivalent to
\bea\label{eone}
\left[\frac{\G'}{\G\R}+\frac{1}{\R^2}\right]
+g^{-1}e^{\beta}\frac{(1-h^2)^{3/2}}{1+3h^2}(\phi')^2
+\frac{3(h')^2}{4(1-h^2)^2}+\frac{3}{4(1-\tfrac{3}{4}\xi^2)^2 }(\xi')^2\nn
+\frac{12e^{\beta}\phi^2\xi^2}{g^2(1-\tfrac{3}{4}\xi^2)^2 }
+\frac{12(-1+h^2+\xi^2)}{g(1-\tfrac{3}{4}\xi^2)^2 (1-h^2)^{3/2}}
=0
\eea
and
\bea\label{etwo}
\frac{\B'}{\R}
+\frac{3}{2}\frac{(h')^2}{(1-h^2)^2}
+\frac{3(\xi')^2}{2(1-\tfrac{3}{4}\xi^2)^2 }+
\frac{24e^{\beta}\phi^2\xi^2}{g^2(1-\tfrac{3}{4}\xi^2)^2}=0
\eea
One can check that the remaining $xx$ component of the Einstein equations,
$E_{xx}=T_{xx}$ with
\bea
\R^{-2}\G^{-1}T_{xx}&=&
g^{-1}e^{\beta}\frac{(1-h^2)^{3/2}}{1+3h^2}(\phi')^2
-\frac{3}{4}\frac{(h')^2}{(1-h^2)^2}
-\frac{3(\xi')^2}{4(1-\tfrac{3}{4}\xi^2)^2}
\nn&&
+\frac{12e^{\beta}\phi^2\xi^2}{g^2(1-\tfrac{3}{4}\xi^2)^2}
-\frac{12(-1+h^2+\xi^2)}{g(1-\tfrac{3}{4}\xi^2)^2(1-h^2)^{3/2}}
\eea
is implied by \reef{eone}, \reef{etwo}.

Observe that the equations \reef{phieqe},
\reef{rhoeqe},
\reef{heqe},
\reef{eone} and \reef{etwo},
can be obtained from the action given in
\reef{actansatz}.

\section{Calculating the on-shell action}
We start by writing \reef{lageinfull} as $S=S_0+S_{CS}$ where
$S_{CS}$ is the metric
independent Chern-Simons like contributions to the action. Since we are
only considering purely electric gauge fields
$S_{CS}$ vanishes on-shell. We further write
$S_0=(16\pi G)^{-1}\int d^4x \sqrt {-g}{\cal L}_0$ and observe that
\be\label{txxlem}
T_{xx}\equiv \tfrac{1}{2}\R^2\left({\cal L}_0-R\right)
\ee
Since $E_{xx}=T_{xx}$ and $E^\mu{}_\mu=-R$ we deduce that
\be
{\cal L}_0=-[E^t{}_t+E^\R{}_\R]
\ee
and hence we can write the on-shell action as
\be
S_{OS}=-\frac{1}{16\pi G}\int dt d^2x dr {\sqrt{-g}} \left(E^t{}_t+E^\R{}_\R\right)
\ee
As in \cite{H32} we next observe that
\be
{\sqrt{-g}} \left(E^t{}_t+E^\R{}_\R\right)=(2\R\G e^{-\B/2})'
\ee
and hence
\bea
S_{OS}&=&-\frac{1}{16\pi G}\int dt d^2 x \int_{r_+}^{\infty}d\R \left[2\R\G e^{-\B/2}\right]'
\eea
An alternative expression for the on-shell action can be obtained by
using the fact that
\be
\sqrt{-g}R=-[r^2e^{-\beta/2}(g'-g\beta')]'-2e^{-\beta/2}gr^2(\frac{g'}{gr}+\frac{1}{r^2})
\ee
Then using \reef{eone} and \reef{txxlem} we obtain
\bea
S_{OS}&=&-\frac{1}{16\pi G}\int dt d^2 x \int_{r_+}^{\infty}d\R
\left[r^2e^{-\beta/2}(g'-g\beta'
-4e^{\beta}\frac{(1-h^2)^{3/2}}{1+3h^2}\phi\phi'
)\right]'\nn
\eea
Finally, we analytically continue by setting $t=-i\tau$ and $iS=-I$, we correspondingly obtain the following two expressions for the on-shell
euclidean action
\bea
I_{OS}&=&\frac{\Delta \tau vol_2}{16\pi G}\int_{r_+}^{\infty}d\R \left[2\R\G e^{-\B/2}\right]'\nn
&=& \frac{\Delta \tau vol_2}{16\pi G}\int_{r_+}^{\infty}d\R \left[r^2e^{-\beta/2}(g'-g\beta'-4e^\beta\frac{(1-h^2)^{3/2}}{1+3h^2}\phi\phi'
)\right]'
\eea
where $\vol_2\equiv \int dx dy$.

\section{Numerical shooting problem}
\label{app:num}

In this paper we have used ODE shooting methods to find the various domain wall and black hole solutions of interest.
We solve the shooting problem by providing data in the IR of the geometry and then integrating into the UV, and also providing data in the UV
of the geometry and
using this to integrate into the IR. We then require that these solutions match correctly at some point in between.
In the UV the data is given by the boundary expansion of perturbations about the skew-whiffed $AdS_4$ solution.
The data in the IR is given by the data at the horizon $r = r_+$ for a black hole solution, or the marginal
and irrelevant perturbations to the $AdS_4$ Pope-Warner and $AdS_2 \times \mathbb{R}^2$ solutions for the zero temperature interpolating solutions
that we construct. Let us denote the IR data as $\{ x_i \}$ and the UV data as $\{ y_a \}$.
We have first order equations for $g, \beta$ and second order equations for $\xi, h, \phi$ and hence if we match the solutions
integrated from the IR up to some intermediate position $r = r_0$ and from the UV down to $r = r_0$, we must match 8 integration
constants associated to these equations. To this end, let us define a function $\Phi^{IR}(x)$ via
\begin{equation}
\Phi^{IR}(x) = \Bigl(g_x(r_0),\, \beta_x(r_0),\, \xi_x(r_0)\,,\xi'_x(r_0)\,, h_x(r_0)\, , h'_x(r_0)\, \phi_x(r_0), \phi'_x(r_0)  \Bigr)\,,
\end{equation}
which is obtained by integrating the differential equations out to a fixed
radius $r_0$ by evolving given IR data $\{ x_i\}$. Here $g_x(r_0)$ denotes the result of integrating the metric function $g(r)$ out to $r=r_0$
subject to IR initial conditions $\{ x_i\}$ and we use analogous notation for the other fields. Let us also define the function
\begin{equation}
\Phi^{UV}(y) = \Bigl(g_y(r_0),\, \beta_y(r_0),\, \xi_y(r_0)\,,\xi'_y(r_0)\,, h_y(r_0)\, , h'_y(r_0)\, \phi_y(r_0), \phi'_y(r_0)  \Bigr)\,,
\end{equation}
which is obtained by evolving given UV initial data $\{ y_a\}$ inward to the same radius $r_0$. We next define a function $V(x,y)$,
which depends on the entire set of initial data
\begin{equation}
V(x,y) = \Phi^{IR}(x) - \Psi^{UV}(y)
\end{equation}
The condition that we have a solution to the differential equations is now simply
\begin{equation}
V(x,y)=0\,,
\end{equation}
which gives rise to 8 conditions. Thus we require that 8 elements of the data $\{ x, y \}$ must be tuned in order to satisfy this.
Let us denote the 8 tuning data variables as $\alpha$, and denote the remaining data that fixes the solution of interest as $\lambda$.
Thus for some $\lambda$ we need to tune the $\alpha$ in order that $V(\alpha ; \lambda) = 0$.

Let us illustrate this using the concrete example of finite temperature black holes.  We parameterize the putative solution by imposing $\xi_1=0$, using the scaling symmetries to fix $\mfac \mu = 1, \beta_a = 0$, and then choosing some fixed $h_1$ in the UV and some fixed $r_+$.  Thus $\lambda = \{ r_+, \mu, \xi_1, \beta_a , h_1 \}$.
We then tune the remaining data, the $\alpha$, which in the IR is $\{ \beta_+, \xi_+, \phi_+,  h_+ \}$ and in the UV is $\{ \varepsilon, \xi_2, h_2, q \}$.
These 8 variables are exactly sufficient degrees of freedom to solve the 8 constraints $V(\alpha ; \lambda) = 0$, and thus one expects that if a solution exists, it is locally unique except at isolated points.

Denote the solution for some $\lambda$ as $\alpha = \alpha_0(\lambda)$. Then we may consider perturbing about that solution $\alpha = \alpha_0 + \delta \alpha$, and the corresponding variation in $V$ is given by $\delta V_i = M_{i j } \delta \alpha_j$ where,
\begin{eqnarray}
M_{ij}(\lambda) =  \left. \partial V_i / \partial \alpha_j \right|_{\alpha_0(\lambda)}
\end{eqnarray}
The local uniqueness of the solution for some $\lambda$ can be expressed as $\det{M_{ij}}(\lambda) \ne 0$ ie. there is no perturbation
to the tuning data $\alpha$ about the solution $\alpha_0$ such that one can maintain $V = 0$.

Whilst generically tuning 8 variables $\alpha$ to fix 8 conditions $V$ to vanish will yield a locally unique solution, it may happen
that at specific $\lambda$ the determinant, $\det{M_{ij}}(\lambda)$ can vanish. In particular, if the original family of solutions
has a different family that branches off at some value of $\lambda$, then this will occur. The zero eigenvector $\delta \alpha$ of $M$
then gives the linear deformation of the solution at $\lambda$ that generates this new branch of solutions.
Considering our example, the unbroken finite temperature black holes have precisely such behaviour at their critical temperature, where
the broken phase solutions emerge as a new branch.

\begin{figure}[ht]
\begin{center}
\includegraphics[width=0.3\textwidth]{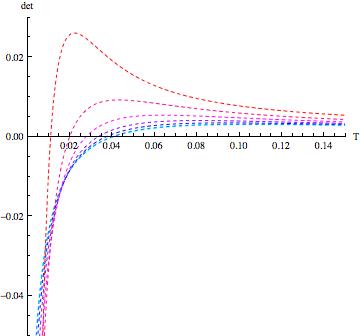}\quad \includegraphics[width=0.3\textwidth]{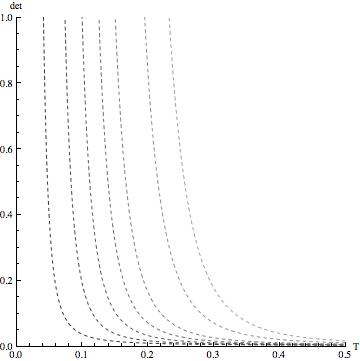}\\
\includegraphics[width=0.3\textwidth]{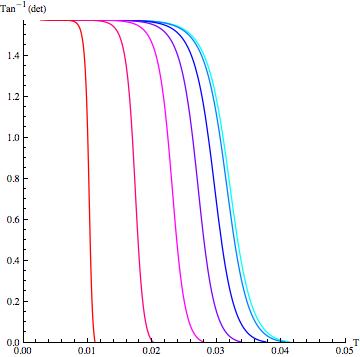}
\caption{Plot showing a quantity $det{M}$ (see text) for the unbroken phase solutions with $h_1<h_1^c$ (top left), the unbroken phase solutions
with $h_1>h_1^c$ (top right) and the broken phase solutions which only exist for $h_1<h_1^c$ (bottom). The
vanishing of $det{M}$ signals a joining point of two solution branches. We see that indeed this quantity vanishes for the critical temperature
of the unbroken solutions with $h_1<h_1^c$, indicating the broken phase emanates from there. For the broken phase, we see that it is correspondingly vanishing
at the beginning of the branches where they join the broken phase. However, for no other values of temperature do we see evidence, within our ansatz,
of new branches of solutions emerging.
\label{det1}}
\end{center}\end{figure}

\begin{figure}[ht]
\begin{center}
\includegraphics[width=0.45\textwidth]{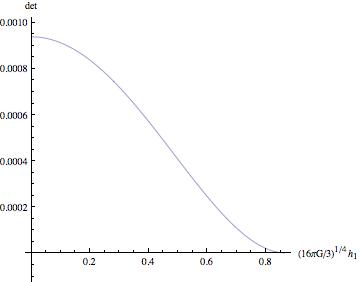}\quad\includegraphics[width=0.45\textwidth]{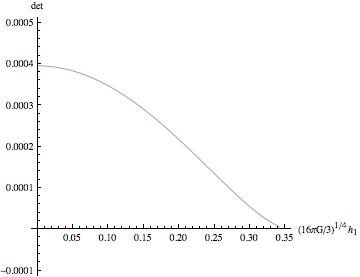}
\caption{Plot showing a quantity $det{M}$ for the uncharged RG domain wall solutions (left) and charged domain wall solutions (right).
Since $det{M}$ appears to be non-vanishing we expect that no other branches of solutions joining these solutions.
\label{det2}}
\end{center}\end{figure}

Hence given our solutions, simply by checking $\det M$ we may numerically check for the existence of marginal deformations that generate
new branches of solutions. We have done so for all the solutions presented in the main text. In Figure \ref{det1} and \ref{det2} we plot
the value of $\det M$ as a function of the various solution parameters for the unbroken and broken finite temperature black holes, and also
for the RG domain walls and the charged skew-whiffed to Pope-Warner and domain wall solutions, respectively.
As expected we see the determinant vanishes for the unbroken
black holes at some critical temperature, for $h_1<h_1^c$, which gives us the superconducting phase transition temperature. This indeed corresponds to the
linear deformation that generates the broken phase branch of solutions. However, apart from this expected branching, we see no evidence of
any other new solution branches emerging from any of the other solutions.

\begin{thebibliography}{99}


\bibitem{Gubser:2008px}
  S.~S.~Gubser,
  ``Breaking an Abelian gauge symmetry near a black hole horizon,''
  Phys.\ Rev.\  D {\bf 78} (2008) 065034
  [arXiv:0801.2977 [hep-th]].

\bibitem{H31}
  S.~A.~Hartnoll, C.~P.~Herzog and G.~T.~Horowitz,
  ``Building a Holographic Superconductor,''
  Phys.\ Rev.\ Lett.\  {\bf 101} (2008) 031601
  [arXiv:0803.3295 [hep-th]].



\bibitem{Hartnoll:2009sz}
  S.~A.~Hartnoll,
  ``Lectures on holographic methods for condensed matter physics,''
  Class.\ Quant.\ Grav.\  {\bf 26} (2009) 224002
  [arXiv:0903.3246 [hep-th]].

\bibitem{Herzog:2009xv}
  C.~P.~Herzog,
  ``Lectures on Holographic Superfluidity and Superconductivity,''
  J.\ Phys.\ A  {\bf 42} (2009) 343001
  [arXiv:0904.1975 [hep-th]].

\bibitem{H32}
  S.~A.~Hartnoll, C.~P.~Herzog and G.~T.~Horowitz,
  ``Holographic Superconductors,''
  JHEP {\bf 0812}, 015 (2008)
  [arXiv:0810.1563 [hep-th]].



\bibitem{dh}
  F.~Denef and S.~A.~Hartnoll,
  ``Landscape of superconducting membranes,''
  Phys.\ Rev.\  D {\bf 79} (2009) 126008
  [arXiv:0901.1160 [hep-th]].

\bibitem{Gauntlett:2009zw}
  J.~P.~Gauntlett, S.~Kim, O.~Varela and D.~Waldram,
  ``Consistent supersymmetric Kaluza--Klein truncations with massive modes,''
  JHEP {\bf 0904} (2009) 102
  [arXiv:0901.0676 [hep-th]].


\bibitem{Gauntlett:2009dn}
  J.~P.~Gauntlett, J.~Sonner and T.~Wiseman,
  ``Holographic superconductivity in M-Theory,''
  Phys.\ Rev.\ Lett.\  {\bf 103} (2009) 151601
  [arXiv:0907.3796 [hep-th]].


\bibitem{Gubser:2009gp}
  S.~S.~Gubser, S.~S.~Pufu and F.~D.~Rocha,
  ``Quantum critical superconductors in string theory and M-theory,''
  arXiv:0908.0011 [hep-th].


\bibitem{Gubser:2009cg}
  S.~S.~Gubser and A.~Nellore,
  ``Ground states of holographic superconductors,''
  Phys.\ Rev.\  D {\bf 80} (2009) 105007
  [arXiv:0908.1972 [hep-th]].

\bibitem{Gubser:2008wz}
  S.~S.~Gubser and F.~D.~Rocha,
  ``The gravity dual to a quantum critical point with spontaneous symmetry
  breaking,''
  Phys.\ Rev.\ Lett.\  {\bf 102}, 061601 (2009)
  [arXiv:0807.1737 [hep-th]].




\bibitem{hr}
  G.~T.~Horowitz and M.~M.~Roberts,
  ``Zero Temperature Limit of Holographic Superconductors,''
  JHEP {\bf 0911} (2009) 015
  [arXiv:0908.3677 [hep-th]].


\bibitem{Gubser:2009qm}
  S.~S.~Gubser, C.~P.~Herzog, S.~S.~Pufu and T.~Tesileanu,
  ``Superconductors from Superstrings,''
  Phys.\ Rev.\ Lett.\  {\bf 103} (2009) 141601
  [arXiv:0907.3510 [hep-th]].

\bibitem{Herzog:2009gd}
  C.~P.~Herzog, I.~R.~Klebanov, S.~S.~Pufu and T.~Tesileanu,
  ``Emergent Quantum Near-Criticality from Baryonic Black Branes,''
  arXiv:0911.0400 [Unknown].


\bibitem{Freund:1980xh}
  P.~G.~O.~Freund and M.~A.~Rubin,
  ``Dynamics Of Dimensional Reduction,''
  Phys.\ Lett.\  B {\bf 97} (1980) 233.


\bibitem{Pope:1984bd}
  C.~N.~Pope and N.~P.~Warner,
  ``An SU(4) Invariant Compactification Of D = 11 Supergravity On A Stretched
  Seven Sphere,''scomb
  Phys.\ Lett.\  B {\bf 150} (1985) 352.

\bibitem{Pope:1984jj}
  C.~N.~Pope and N.~P.~Warner,
  ``Two New Classes Of Compactifications Of D = 11 Supergravity,''
  Class.\ Quant.\ Grav.\  {\bf 2} (1985) L1.

\bibitem{Englert:1982vs}
  F.~Englert,
  ``Spontaneous Compactification Of Eleven-Dimensional Supergravity,''
  Phys.\ Lett.\  B {\bf 119} (1982) 339.

\bibitem{Awada:1982pk}
  M.~A.~Awada, M.~J.~Duff and C.~N.~Pope,
  ``N = 8 supergravity breaks down to N = 1,''
  Phys.\ Rev.\ Lett.\  {\bf 50}, 294 (1983).


\bibitem{Duff:1984sv}
  M.~J.~Duff, B.~E.~W.~Nilsson and C.~N.~Pope,
  ``The Criterion For Vacuum Stability In Kaluza-Klein Supergravity,''
  Phys.\ Lett.\  B {\bf 139} (1984) 154.

\bibitem{Berkooz:1998qp}
  M.~Berkooz and S.~J.~Rey,
  ``Non-supersymmetric stable vacua of M-theory,''
  JHEP {\bf 9901} (1999) 014
  [Phys.\ Lett.\  B {\bf 449} (1999) 68]
  [arXiv:hep-th/9807200].

\bibitem{am}
A.~Murugan
``Renormalisation group flows in gauge-gravity duality",
PhD Thesis, Princeton University, September 2009.

\bibitem{Page:1984fu}
  D.~N.~Page and C.~N.~Pope,
  ``Instabilities In Englert Type Supergravity Solutions,''
  Phys.\ Lett.\  B {\bf 145} (1984) 333.

\bibitem{Gubser:2009qt}
  S.~S.~Gubser and F.~D.~Rocha,
  ``Peculiar properties of a charged dilatonic black hole in $AdS_5$,''
  arXiv:0911.2898 [Unknown].


\bibitem{Goldstein:2009cv}
  K.~Goldstein, S.~Kachru, S.~Prakash and S.~P.~Trivedi,
  ``Holography of Charged Dilaton Black Holes,''
  arXiv:0911.3586 [hep-th].



\bibitem{Buchel:2006gb}
  A.~Buchel and J.~T.~Liu,
  ``Gauged supergravity from type IIB string theory on Y(p,q) manifolds,''
  Nucl.\ Phys.\  B {\bf 771} (2007) 93
  [arXiv:hep-th/0608002].



%
\bibitem{Gauntlett:2007ma}
  J.~P.~Gauntlett and O.~Varela,
  ``Consistent Kaluza-Klein Reductions for General Supersymmetric AdS
  Solutions,''
  Phys.\ Rev.\  D {\bf 76} (2007) 126007
  [arXiv:0707.2315 [hep-th]].

\bibitem{gp}
  J.~P.~Gauntlett and S.~Pakis,
  ``The geometry of D = 11 Killing spinors,''
  JHEP {\bf 0304} (2003) 039
  [arXiv:hep-th/0212008].

\bibitem{Englert:1983qe}
  F.~Englert, M.~Rooman and P.~Spindel,
  ``Supersymmetry Breaking By Torsion And The Ricci Flat Squashed Seven
  Spheres,''
  Phys.\ Lett.\  B {\bf 127} (1983) 47.

\bibitem{Kovtun:2008kw}
  P.~Kovtun and A.~Ritz,
  ``Black holes and universality classes of critical points,''
  Phys.\ Rev.\ Lett.\  {\bf 100} (2008) 171606
  [arXiv:0801.2785 [hep-th]].

\bibitem{Franco:2009yz}
  S.~Franco, A.~Garcia-Garcia and D.~Rodriguez-Gomez,
  ``A general class of holographic superconductors,''
  arXiv:0906.1214 [hep-th].










\bibitem{Chamblin:1999tk}
  A.~Chamblin, R.~Emparan, C.~V.~Johnson and R.~C.~Myers,
  ``Charged AdS black holes and catastrophic holography,''
  Phys.\ Rev.\  D {\bf 60} (1999) 064018
  [arXiv:hep-th/9902170].


\bibitem{Cvetic:1999ne}
  M.~Cvetic and S.~S.~Gubser,
  ``Phases of R-charged black holes, spinning branes and strongly coupled
  gauge theories,''
  JHEP {\bf 9904} (1999) 024
  [arXiv:hep-th/9902195].





\bibitem{bk}
  V.~Balasubramanian and P.~Kraus,
  ``A stress tensor for anti-de Sitter gravity,''
  Commun.\ Math.\ Phys.\  {\bf 208} (1999) 413
  [arXiv:hep-th/9902121].

\bibitem{st}
  K.~Skenderis and P.~K.~Townsend,
  ``Gravitational stability and renormalization-group flow,''
  Phys.\ Lett.\  B {\bf 468} (1999) 46
  [arXiv:hep-th/9909070].



%
\bibitem{Hartnoll:2007ip}
  S.~A.~Hartnoll and C.~P.~Herzog,
  ``Ohm's Law at strong coupling: S duality and the cyclotron resonance,''
  Phys.\ Rev.\  D {\bf 76} (2007) 106012
  [arXiv:0706.3228 [hep-th]].




\end{thebibliography}
\end{document}